\documentclass[11pt,reqno,english]{amsart}
\usepackage{amsfonts,amsmath,latexsym,verbatim,amscd,mathrsfs,color,array,amssymb,amsthm,graphicx,mathtools}
\usepackage[normalem]{ulem}
\usepackage[overload]{empheq}   

\usepackage{enumerate}
\usepackage[colorlinks=true]{hyperref}
\usepackage[hmargin=2.5cm, vmargin=2.5cm]{geometry}	
\usepackage{pdfsync}
\usepackage{epstopdf}
\usepackage{diagbox}
\usepackage{subcaption}

\usepackage{multirow}

\usepackage[justification=centering]{caption}

\usepackage{amsmath,amssymb,amsthm,amsfonts,graphicx,color}
\usepackage{amssymb}

\usepackage{amsfonts,amsthm,amsmath,latexsym,bm,graphics}

\newcommand{\ep}{\varepsilon}
\newcommand{\Ro}{\mathcal{R}_0}
\newcommand{\erfc}{\mathrm{erfc}}

\newcommand{\asy}{\mathrm{asy}}
\newcommand{\tot}{\mathrm{tot}}
\newcommand{\dimvar}[1]{\boldsymbol{\mathrm{#1}}}

\theoremstyle{rem}
\newtheorem{rem}{rem}

\begin{document}

\title[Prevalence and the rate of antigenic evolution]
      {Prevalence and the rate of antigenic evolution are jointly determined in an eco-evolutionary SIR model}

\author{Daniel Gomez}
\address{Department of Mathematics and Statistics, University of New Mexico, Albuquerque, NM 87131, USA}
\email{danielgomez@unm.edu}

\author{Joshua B. Plotkin}
\address{Department of Biology, University of Pennsylvania, Philadelphia, PA 19104, USA}
\email{jplotkin@sas.upenn.edu}


\date{\today}

\begin{abstract}\hspace{2ex}\\
\textbf{Abstract:} Influenza A persists by continually changing the antigens that its host population recognizes. How fast it evolves and how many hosts it infects are not independent, because the immunity that selects new variants is produced by the infections the pathogen has already caused. We analyze a model of Andreasen et.\ al.\ that captures this eco-evolutionary feedback, and we extend it to the case where infection may kill the host. In the model infected and recovered hosts are distributions over a one-dimensional antigenic space coupled by cross-immunity. The endemic state is a traveling pulse: the infected distribution keeps a fixed shape and advances through antigenic space at constant speed. In the limit of weak antigenic mutation we construct the pulse asymptotically and find that: (i) the speed of antigenic evolution decreases as prevalence rises, (ii) less standing antigenic variation accompanies either faster or slower antigenic evolution, depending on whether pathogens differ in immune evasion or host lifespan  (iii) the total number of infected hosts is largest at an intermediate virulence when the host birth rate is sufficiently low.

\noindent\textbf{Relevance to Life Science}Antigenic drift of influenza A is the continual replacement of circulating variants by successors that host immunity no longer recognizes. The epidemic and the evolution occur on the same time scale and cannot be studied separately: each new variant is selected by immunity that the epidemic itself produces. We find that more infection means slower antigenic evolution, not faster, because infected hosts are unavailable to new variants. Influenza A in wild waterfowl illustrates this: infection occupies a far larger fraction of a duck's life than of a human's, so the model predicts higher prevalence, slower antigenic evolution, and greater standing antigenic variance in the avian reservoir, consistent with the observed stasis of avian haemagglutinin. Our results also show that the relation between variance and rate can have the opposite sign: measles viruses produce far less immune evasion per infection than influenza, and the model predicts both less antigenic variance and slower evolution. For a pathogen that can kill its host, we find that the number of infections is largest at an intermediate virulence when births are too slow to replace the hosts virulence removes. We show how two classical results of evolutionary theory hold exactly at the endemic state, with antigenic position as the trait and a variant's growth rate as its fitness. By the Price equation the pathogen advances through antigenic space at a rate equal to the covariance between growth rate and antigenic position. By Fisher's fundamental theorem, with the deterioration of the environment produced by the pathogen itself, selection raises the mean growth rate at exactly the rate at which immunity from past infections lowers it.

\noindent\textbf{Mathematical Content:} We investigate in detail the structure of traveling pulse solutions to a system of nonlocal partial differential equations. This system constitutes an extension of the classical SIR framework in which the infected and recovered populations now occupy a one-dimensional antigenic space. Antigenic drift is described by classical diffusion within the infected population while re-infection is described by a convolution over antigenic space between the recovered distribution and a cross-immunity kernel. The pulse speed is selected at its leading edge. In the limit of weak antigenic drift a WKB analysis shows that the infected distribution is Gaussian to leading order and hence set by two scalars, the prevalence and the antigenic variance, for which closure conditions give two nonlinear equations. With this framework we perform a comprehensive numerical study across broad parameter regimes, supplemented with further analytical approximations in the regimes where infections are either short- or long-lived. Our asymptotic results are validated through full numerical simulations of the original system of partial differential equations at select parameter values.
\end{abstract}
\maketitle


\section{Introduction}\label{sec:intro}

Influenza A persists in the human population by escaping the immunity it induces. Neutralizing antibodies bind principally to the globular head of the haemagglutinin glycoprotein, and amino acid substitutions in its HA1 domain can abolish that binding. The HA1 domain accordingly evolves faster than the rest of the genome, accumulating roughly six nucleotide substitutions per genome per year against two in the non-structural gene \cite{lin_2003}. Each substitution that alters an epitope allows the virus to infect hosts who were immune to its predecessors. The result is antigenic drift, a continual turnover of circulating variants in which each is displaced by a successor that the standing immunity of the host population no longer recognizes \cite{nelson_2007}.

The genealogy that antigenic drift produces has a distinctive shape. For human influenza A/H3N2, sampled haemagglutinin lineages coalesce onto a single trunk, and side branches persist for only a few years before going extinct \cite{nelson_2007,bedford_2014}. Standing antigenic diversity at any moment is correspondingly low, and sequence data resolve that diversity into a succession of discrete clusters rather than a diffuse cloud \cite{plotkin_2002}. Antigenic cartography applied to haemagglutination inhibition assays gives the same picture from phenotype rather than sequence: strains fall into a succession of antigenic clusters, each replacing the last, ordered along a single dominant direction \cite{smith_2004,koelle_2006}. That the diversity remains so low is itself a fact requiring explanation, and models invoking short-lived strain-transcending immunity \cite{ferguson_2003} or discrete structuring of the strain space by cross-immunity \cite{gupta_1998} have been advanced to account for it. What the trunk-dominated genealogy licenses, and what we use it for here, is the idealization adopted by Lin et al.\@ \cite{lin_2003} in which antigenic space is one-dimensional and the virus advances along a single axis of escape. The idealization is a strong one. Haemagglutinin carries several antigenic sites that can change independently, so escape is not confined to one direction, and reassortment between co-circulating lineages produces antigenic change that no diffusion process describes \cite{nelson_2007}. We return to both limitations in \S\ref{sec:discussion}.

Antigenic drift cannot be separated from the epidemiology that drives it, and the model of Lin et al.\@ \cite{lin_2003} that we take up here is eco-evolutionary in a strong and literal sense. The ecology is the infection dynamics: the flow of hosts between the susceptible, infected, and recovered classes, and the size of the host population itself. The evolution is the traveling wave: the advance of the pathogen through antigenic space. The ecology and the evolution are not two models joined through a shared parameter. They are one system of equations, and neither half can be solved without the other. The speed of the wave depends on the prevalence of infection, because prevalence determines how many hosts remain available to the variants at the front. The prevalence depends in turn on where the wave sits relative to the immunity the host population has already accumulated, since that immunity is what limits transmission. Evolutionary rate and epidemiological state are thus determined jointly, and the endemic equilibrium of the epidemiology is simultaneously the steady advance of the evolution.

Pease \cite{pease_1987} identified the mechanism behind this coupling. In the classical theory of a non-evolving childhood infection such as measles, the susceptible class is replenished by demographic turnover: immune hosts die and susceptible hosts are born. An antigenically evolving pathogen has a second and typically faster source of susceptible hosts. The pathogen changes, and hosts immune to its predecessors become susceptible again without any host having been born or having died. Antigenic evolution is therefore not a process running alongside the epidemiology. Antigenic evolution is one of the terms sustaining the epidemic, and the dominant one whenever infections are short relative to the host lifespan.

Antigenically evolving pathogens are worth studying as a class for this reason, and nonlocal partial differential equations in antigenic space are the natural setting. The formulation carries two distributions over the same axis, the pathogen's occupancy of antigenic space and the host population's immunity to it, and lets them interact nonlocally through a cross-immunity kernel. Compartmental models with a fixed set of strains can represent the ecology but must specify the evolution in advance; models of adaptation in a fitness landscape can represent the evolution but hold the ecology fixed as a background. The continuum formulation represents both.

Traveling waves arise in two distinct traditions in evolutionary and ecological theory. In the first the wave moves through physical space, as when a gene or an epidemic spreads geographically \cite{aronson_1978}; recent work in this tradition asks how selection on a trait such as virulence differs at the front of a spreading epidemic from selection behind it \cite{griette_2015}. In the second the wave moves through a space of genotypes or phenotypes, and its speed equals the rate of adaptation. Tsimring et al.\@ \cite{tsimring_1996} introduced models of this second kind for RNA viruses. Subsequent work developed them into a theory of adaptation in asexual populations, in which the population occupies a localized distribution in fitness whose mean advances at a speed determined by the mutational input and the strength of selection \cite{desai_2007,rouzine_2008,hallatschek_2011}. Our model belongs to the second tradition, as do the antecedent formulations of antigenic drift as a continuous process in a strain space \cite{pease_1987,sasaki_1994,andreasen_1996}. Our model differs from the population-genetic versions in one respect. Those specify the fitness gradient externally and hold it fixed, whereas here the gradient depends on the recovered distribution, which the pathogen itself produces. The speed therefore cannot be computed without simultaneously solving for the epidemiological state.

Leveraging the trunk-dominated genealogy of influenza A, the model of Andreasen et al.\@ \cite{andreasen_1996} collapses antigenic space into the one-dimensional real line. The time evolution of distributions of infected and recovered populations over antigenic space are described by an SIR-type system of nonlocal partial differential equations. Antigenic drift is modeled by diffusion in antigenic space within the infected population, while reinfections are modeled by a convolution between the recovered distribution and a cross-immunity kernel. The authors incorporate evidence of lifelong cross-immunity to influenza A \cite{couch_1983,potter_1977} by restricting the functional form of the cross-immunity kernel so that recovery from a strain $\dimvar{x}\in\mathbb{R}$ grants lifelong immunity to all strains $\dimvar{y}\leq\dimvar{x}$. On the other hand, immune evasion resulting from epitope alterations is incorporated into the cross-immunity kernel so that immunity to strain $\dimvar{y}>\dimvar{x}$ decreases as $\dimvar{y}$ increases. A more detailed description of the model is found in \S\ref{subsec:model-description} below.

One of the central features of the model of Andreasen et al.\@ is that it exhibits traveling pulse solutions. In a subsequent extension by Lin et al.\@ \cite{lin_2003} the authors determined  the speed of these pulse solutions and furthermore used asymptotic analysis to approximate the profiles of both the infected and recovered population distributions over antigenic space. Despite its simplicity, this modified SIR-type model serves as a promising, yet relatively unexplored, framework in which to investigate the eco-evolutionary dynamics of infectious disease. Our objective in revisiting this model is twofold, the first of which is to introduce infection-induced mortality, or \textit{virulence}, into the model and characterize its consequences on the eco-evolutionary dynamics of the system. 

Why virulence should take any particular value is an old question. The view that prevailed into the middle of the twentieth century held that a well-adapted parasite evolves toward avirulence, on the reasoning that killing the host destroys the resource the parasite depends on. Levin and Pimentel \cite{levin_1981}, and then Anderson and May \cite{anderson_1982}, replaced this argument with one from fitness. For a parasite spreading in a host population with a susceptible density $\dimvar{S}$, the basic reproduction number is
\begin{equation}\label{eq:classical-R0}
    \Ro = \frac{\dimvar{\beta}\,\dimvar{S}}{\dimvar{\mu} + \dimvar{\nu} + \dimvar{\omega}},
\end{equation}
where $\dimvar{\beta}$ is the transmission rate, $\dimvar{\mu}$ the background host mortality, $\dimvar{\nu}$ the recovery rate, and $\dimvar{\omega}$ the infection-induced mortality, that is, the virulence. Virulence appears in the denominator because a parasite that kills its host ends its own infectious period, so selection acts against virulence directly. Were there no compensating benefit, the prediction would indeed be avirulence. The trade-off hypothesis supposes instead that transmission and virulence are physiologically linked, $\dimvar{\beta}=\dimvar{\beta}(\dimvar{\omega})$, and that selection maximizes $\Ro$ over $\dimvar{\omega}$. An interior optimum exists precisely when transmission increases at a diminishing rate, $\dimvar{\beta}''(\dimvar{\omega})<0$ \cite{alizon_2009,day_2007}. Intermediate virulence is therefore a consequence of an assumed constraint, together with the assumption that the pathogen sits at a single-strain endemic equilibrium. We note that the denominator of \eqref{eq:classical-R0} is exactly the time scale $\dimvar{\tau}$ by which we non-dimensionalize in \S\ref{subsec:non-dimensional}, so that the parameter $\Ro$ of our model is the classical basic reproduction number of \eqref{eq:classical-R0} evaluated in a fully susceptible population.

Subsequent theories of virulence evolution relaxed both assumptions. When hosts carry more than one strain, within-host competition rewards faster exploitation and selects for higher virulence than $\Ro$ maximization predicts \cite{van_1995,alizon_2013,alizon_2008decreased}, and the operative trade-off may involve recovery rather than mortality \cite{alizon_2008transmission}. Day and Gandon \cite{day_2007,day_2012} showed that an invasion analysis, which assumes epidemiological dynamics are fast relative to evolutionary ones, discards information that a population-genetic formulation retains. Writing the change in mean virulence as a Price equation, they obtain
\begin{equation}\label{eq:day-gandon-price}
    \frac{d\overline{\dimvar{\omega}}}{dt} = \dimvar{S}\,\sigma_{\dimvar{\omega}\dimvar{\beta}} - \sigma_{\dimvar{\omega}\dimvar{\omega}},
\end{equation}
in which selection against virulence acts with constant strength while selection for transmission acts with strength proportional to the density of susceptible hosts. The rate of evolution is then a product of a variance and a selection gradient, and transient change can run opposite to the long-term outcome \cite{day_2007}. Griette et al.\@ \cite{griette_2015} showed that virulence at the front of a spatially spreading epidemic differs from virulence in the bulk, for the same reason: the front is not at equilibrium.

Two features of this classical account do not carry over to the model considered here. First, we impose no trade-off: the basic reproduction number and the virulence enter our non-dimensional model as independent parameters, and transmission is not a function of virulence. Second, the pathogen is not at a single-strain endemic equilibrium. A continuum of antigenic variants circulates, the recovered class is continually reinfected, and the state we analyze is a traveling pulse rather than a fixed point. Neither the argument for avirulence nor the argument for an interior optimum through $\dimvar{\beta}''(\dimvar{\omega})<0$ applies. We nonetheless find in \S\ref{subsec:optimal-virulence} that intermediate virulence can maximize the total number of infections, and for a different reason, namely that virulence depresses the equilibrium host population size.

We treat virulence as a fixed parameter of the pathogen, not as an evolving trait. Our antigenic variable $\dimvar{x}$ evolves; $\dimvar{\omega}$ does not. The results below therefore describe how a given level of virulence shapes antigenic evolution and prevalence, and which level of virulence would maximize the number of infections. They are not statements about the level of virulence that selection produces. We discuss in \S\ref{sec:discussion} what would be required to let antigenic type and virulence evolve jointly, which is the natural next step and the point at which the population-genetic formulation of \cite{day_2007,day_2012} becomes directly applicable.

The second objective in revisiting the model of Lin et al.\@ is to determine how variations in demographic and epidemiological parameters affect infection outcomes. In particular, we are interested in considering parameter regimes beyond those relevant for influenza A in human populations, and therefore falling outside the range of the parameter values considered by Lin et al.\@ \cite{lin_2003}.  Two dimensionless parameters carry most of the biological interpretation. The first, $m$, is the fraction of infections that end in death from causes unrelated to the infection rather than in recovery; equivalently, it measures the duration of infection relative to the lifespan of the host. The second, $\ep$, measures the immune evasion a lineage accumulates over the course of one average infection, and so plays the role of mutation. Our results separate the effects of these two parameters, which tend to influence the speed of antigenic evolution and the standing antigenic variance in opposite ways. Holding $\ep$ fixed, longer-lived infections give slower evolution together with greater antigenic variance. Varying $\ep$, a pathogen that evades immunity more slowly has both slower evolution and lower antigenic variance. Comparisons between real pathogens therefore require care about which parameter distinguishes them, a point we develop in \S\ref{sec:discussion} using influenza A in humans, influenza A in wild waterfowl, and measles.

Models coupling epidemiology to evolution are usually studied either by invasion analysis, which assumes the epidemiology equilibrates fast enough to be treated as a constraint, or by direct simulation. The first assumption is untenable here by construction, since the antigenic and epidemiological dynamics proceed on a common time scale \cite{day_2007}; and simulation yields the solution at particular parameter values without exhibiting the dependence on parameters. We therefore study the traveling pulse as a solution of a nonlinear, nonlocal partial differential equation, and we construct it asymptotically in the limit of small antigenic diffusion (mutation). The infinite-dimensional problem then collapses onto two scalars, the prevalence $I_\asy$ and the antigenic variance $\sigma_\asy^2$, which satisfy a pair of algebraic equations. The profiles of the infected and recovered distributions, and the speed at which the pathogen advances through antigenic space, follow from these two quantities. This is what makes systematic comparison across parameters possible, and it allows us to ask how the rate of antigenic evolution differs between host species or between pathogens.

We highlight one feature of the construction at the outset. The leading edge of the pulse, where infections are rare, selects the speed, in the manner familiar from pulled fronts \cite{aronson_1978}. Speeds determined this way are known to be sensitive to the far tail of the distribution, and in a finite population the discreteness of individuals cuts that tail off and reduces the speed by a term decaying only logarithmically in population size \cite{brunet_1997}. Lin et al.\@ \cite{lin_2003} estimated this correction for influenza and found it small but not negligible. Our results are deterministic throughout, and we return to this point in \S\ref{sec:discussion}.

The remainder of the paper is organized as follows. In \S\ref{subsec:model-description} and \S\ref{subsec:non-dimensional} we introduce in more detail the SIR model with virulence and its non-dimensional form. The bulk of our asymptotic analysis is found in  \S\ref{sec:traveling-pulse} where we first use a WKB-type analysis to reduce the calculation of traveling pulse solutions to solving a system of two nonlinear equations for the total proportion of infected individuals and its variance. The biological relevance of the two nonlinear equations is established in  \S\ref{subsec:identities} where we identify their relationship to the Price equation and Fisher's fundamental theorem of natural selection. In \S\ref{subsec:m-regimes} we identify two asymptotic regimes, corresponding to long-lasting infections ($m\gg\sqrt{\ep}$) and to short-lasting infections ($\ep^2\ll m\ll\ep$), for which the nonlinear system is further simplified, and in \S\ref{subsec:exponential-kernel} we introduce an exponential cross-immunity kernel which provides computational simplifications. In \S\ref{sec:results} we collect numerical results showcasing different infection outcomes as parameter values are varied. In addition in \S\ref{subsec:optimal-virulence} we illustrate how intermediate virulence maximizes the endemic burden of infection. Finally, in \S\ref{sec:discussion} we summarize our key results, discuss their biological significance, outline some of the shortcomings of the model, and suggest directions for future work.

\subsection{The Modified SIR Model}\label{subsec:model-description}

The model proposed by Lin et al.\@ \cite{lin_2003} extends the classical SIR compartmental model by introducing antigenic drift and cross-immunity on a one-dimensional continuous antigenic space $-\infty<\dimvar{x}<+\infty$. The susceptible population $\dimvar{S}(\dimvar{t})$ depends only on time whereas the infected and recovered populations are distributed over antigenic space and are respectively denoted by $\dimvar{i}(\dimvar{x},\dimvar{t})$ and $\dimvar{r}(\dimvar{x},\dimvar{t})$. Antigenic drift is modeled by diffusion in $\dimvar{x}$ within the infected population while cross-immunity is modeled by a \textit{cross-immunity kernel} $\dimvar{K}(\dimvar{z})$ that mediates interactions between the infected and recovered populations. Specifically, in this model the growth rate of the infected population due to its interactions with the recovered and susceptible populations is given by
\begin{equation}\label{eq:infected-growth-rate}
    \dimvar{\beta}(\dimvar{N}(\dimvar{t}))\int_{-\infty}^{\infty}\dimvar{i}(\dimvar{x},\dimvar{t})\dimvar{K}(\dimvar{x}-\dimvar{y})\dimvar{r}(\dimvar{y},\dimvar{t})d\dimvar{y}  + \dimvar{\beta}(\dimvar{N}(\dimvar{t}))\dimvar{i}(\dimvar{x},\dimvar{t})\dimvar{S}(\dimvar{t}),
\end{equation}
where $\dimvar{\beta}(\cdot)$ is a contact rate per capita dependent on the total population size
\begin{subequations}
\begin{equation}\label{eq:dim-N-tot}
	\dimvar{N}(\dimvar{t}) := \dimvar{S}(\dimvar{t}) +  \dimvar{I}(\dimvar{t}) +  \dimvar{R}(\dimvar{t}),
\end{equation}
and where
\begin{equation}
	\dimvar{I}(\dimvar{t}) := \int_{-\infty}^\infty \dimvar{i}(\dimvar{x},\dimvar{t})d\dimvar{x},\qquad \dimvar{R}(\dimvar{t}) := \int_{-\infty}^\infty \dimvar{r}(\dimvar{x},\dimvar{t})d\dimvar{x}.
\end{equation}
\end{subequations}
The convolution between $\dimvar{r}(\cdot,\dimvar{t})$ and $\dimvar{K}(\cdot)$  in \eqref{eq:infected-growth-rate} describes the likelihood that individuals recovered from strain $\dimvar{y}$ are infected by strain $\dimvar{x}$. Two key assumptions, namely \textit{lifelong immunity} to \textit{past} strains and \textit{cross-immunity} to \textit{future} strains, are modeled by assuming that
\begin{enumerate}
    \item[i.)] $\dimvar{K}(\dimvar{z})=0$ for all $z\leq 0$, and
    \item[ii.)] $\dimvar{K}(\dimvar{z})$ is monotone increasing and $0<\dimvar{K}(\dimvar{z})\leq 1$ for all $\dimvar{z}>0$.
\end{enumerate}
Both assumptions have direct empirical support for influenza A. Homotypic immunity, meaning resistance to reinfection by the same virus, is potent and of long duration \cite{couch_1983}, which is the content of assumption i.). Protection against an antigenically distinct strain of the same subtype is graded rather than all-or-nothing, and is ordered by antigenic distance: in challenge experiments the proportion of immunized volunteers who could subsequently be infected increased with the antigenic distance between the immunizing strain and the challenge virus \cite{potter_1977}. Assumption ii.) is the idealization of this observation, with $\dimvar{K}(\dimvar{z})$ read as the susceptibility of a host recovered from a strain at antigenic distance $\dimvar{z}$ in the past.
Finally, we make the technical assumption that 
\begin{enumerate}
    \item[iii.)]  $0<\dimvar{a}:=\lim_{z\rightarrow 0^+} \dimvar{K}(\dimvar{z})/\dimvar{z} < +\infty$,
\end{enumerate}
which describes a linear decrease in immunity to neighboring future strains. With these assumptions on the cross-immunity kernel the one-dimensional antigenic space inherits a directionality in the sense that if $\dimvar{x}>\dimvar{y}$ then $\dimvar{x}$ is considered a \textit{future} strain and $\dimvar{y}$ is a \textit{past} strain. 

With the above considerations the eco-evolutionary SIR model takes the form
\begin{subequations}\label{eq:dim-sys}
\begin{gather}
    \frac{d\dimvar{S}(\dimvar{t})}{d\dimvar{t}} = \dimvar{N}(\dimvar{t})\dimvar{b}(\dimvar{N}(\dimvar{t}))  - \dimvar{\mu}\dimvar{S}(\dimvar{t}) - \dimvar{\beta}(\dimvar{N}(\dimvar{t}))\dimvar{S}(\dimvar{t})\int_{-\infty}^\infty \dimvar{i}(\dimvar{x},\dimvar{t})d\dimvar{x},\hspace{22ex} \label{eq:dim-sys-S}\\
    \begin{split}\frac{\partial \dimvar{i}(\dimvar{x},\dimvar{t})}{\partial \dimvar{t}} = \dimvar{D}\frac{\partial^2\dimvar{i}(\dimvar{x},\dimvar{t})}{\partial\dimvar{x}^2} +  \dimvar{\beta}(\dimvar{N}(\dimvar{t}))\dimvar{i}(\dimvar{x},\dimvar{t})\biggl(\int_{-\infty}^{\dimvar{x}} \dimvar{K}(\dimvar{x}-\dimvar{y})\dimvar{r}(\dimvar{y},\dimvar{t})& d\dimvar{y}  + \dimvar{S}(\dimvar{t})\biggr) \\
    & - (\dimvar{\mu} + \dimvar{\nu} + \dimvar{\omega}) \dimvar{i}(\dimvar{x},\dimvar{t}),\end{split}\label{eq:dim-sys-i} \\
    \frac{\partial\dimvar{r}(\dimvar{x},\dimvar{t})}{\partial\dimvar{t}} = -\dimvar{\beta}(\dimvar{N}(\dimvar{t}))\dimvar{r}(\dimvar{x},\dimvar{t})\int_{\dimvar{x}}^\infty \dimvar{K}(\dimvar{y}-\dimvar{x})\dimvar{i}(\dimvar{y},\dimvar{t})d\dimvar{y} + \dimvar{\nu}\dimvar{i}(\dimvar{x},\dimvar{t}) - \dimvar{\mu} \dimvar{r}(\dimvar{x},\dimvar{t}), \hspace{11ex}\label{eq:dim-sys-r}
\end{gather}
\end{subequations}
where $\dimvar{\mu}$ is the base death rate, $\dimvar{\nu}$ is the recovery rate, $\dimvar{\omega}$ is the infection-induced death rate, and $\dimvar{b}(\cdot)$ is the birth rate depending on the total population size. We will be interested in providing a comprehensive characterization of endemic equilibrium solutions to \eqref{eq:dim-sys} in the form of traveling pulse solutions. In particular, through an asymptotic analysis of \eqref{eq:dim-sys} in the limit of small antigenic drift we determine how the prevalence of infection is influenced by both epidemiological and demographic parameters. This analysis extends the original work of Lin et al.\@ \cite{lin_2003} in which parameters were chosen to reflect influenza A in humans and infection-induced death was neglected, i.e.\@ $\dimvar{\omega}=0$. Differentiating \eqref{eq:dim-N-tot} and using \eqref{eq:dim-sys} gives 
\begin{equation}\label{eq:dim-sys-N}
	\frac{d\dimvar{N}(\dimvar{t})}{d\dimvar{t}} = \dimvar{N}(\dimvar{t})\dimvar{b}(\dimvar{N}(\dimvar{t}))  - \dimvar{\mu} \dimvar{N}(\dimvar{t}) - \dimvar{\omega} \dimvar{I},
\end{equation}
from which we see that in the absence of infection-induced fatalities the total population size in an endemic equilibrium depends only on demographic parameters. However, if $\dimvar{\omega}>0$ then this total population size depends non-trivially on epidemiological parameters through the solution of \eqref{eq:dim-sys}. One of the outcomes of our asymptotic analysis is the determination of this non-trivial dependence.

Finally, we conclude by making concrete choices for the birth rate $\dimvar{b}(\cdot)$ and contact rate per capita $\dimvar{\beta}(\cdot)$. Specifically we assume \textit{logistic growth} and \textit{standard incidence} so that
\begin{equation}
	\dimvar{b}(\dimvar{N}) = \dimvar{b}_0\left(1 - \frac{\dimvar{N}}{Q}\right), \qquad \dimvar{\beta}(\dimvar{N}) = \frac{\dimvar{\beta}_0}{\dimvar{N}},
\end{equation}
where $\dimvar{b}_0$ is the intrinsic birth rate, $Q$ is the carrying capacity, and $\dimvar{\beta}_0$ is the contact rate. We remark that while \textit{mass action} is a more appropriate choice for airborne illnesses, our overall results in the subsequent sections are similar for both mass-action and standard incidence provided that infection-induced mortality is small or absent. Therefore, apart from our discussion on the effects of infection-induced mortality in \S\ref{subsec:optimal-virulence}, if $\dimvar{\omega}=0$ then our results are equally applicable for both standard and mass-action incidence rates.

\subsection{Non-Dimensional Form of the Modified SIR Model}\label{subsec:non-dimensional}

We non-dimensionalize the modified SIR model \eqref{eq:dim-sys} by letting
\begin{subequations}\label{eq:non-dim-variables}
\begin{equation}
	t = \dimvar{\tau}\dimvar{t}\quad\text{and}\quad x = \dimvar{a}\,\dimvar{x},\quad\text{where}\quad \dimvar{\tau} := \dimvar{\mu}+\dimvar{\nu}+\dimvar{\omega},
\end{equation}
where we remind the reader that $\dimvar{a}= \lim_{z\rightarrow 0^+}\dimvar{K}(\dimvar{z})/\dimvar{z}$ and remark that $\dimvar{\tau}^{-1}$ reflects the average infection duration. Note that with this rescaling time is normalized with respect to the average infection duration while antigenic space is rescaled so that cross-immunity has a growth rate of unity at the origin. In addition we let $N(t) = \dimvar{N}(t/\dimvar{\tau})$ in terms of which we define
\begin{equation}
	S(t) := \frac{1}{N(t)}\dimvar{S}\left(\frac{t}{\dimvar{\tau}}\right),\quad i(x,t) := \frac{1}{\dimvar{a}N(t)}\dimvar{i}\left(\frac{x}{\dimvar{a}},\frac{t}{\dimvar{\tau}}\right),\quad r(x,t):= \frac{1}{\dimvar{a}N(t)} \dimvar{r}\left(\frac{x}{\dimvar{a}},\frac{t}{\dimvar{\tau}}\right),
\end{equation}
respectively describing the \textit{proportion} of the susceptible population, and of the infected and recovered antigenic distributions. Note that
\begin{equation}
    I(t) := \int_{-\infty}^\infty i(x,t)dx = \frac{\dimvar{I}(t/\dimvar{\tau})}{\dimvar{N}(t/\dimvar{\tau})}\quad\text{and}\quad R(t) := \int_{-\infty}^\infty r(x,t)dx = \frac{\dimvar{R}(t/\dimvar{\tau})}{\dimvar{N}(t/\dimvar{\tau})}
\end{equation}
\end{subequations}
respectively denote the total proportion of the infected and recovered populations. Substituting \eqref{eq:non-dim-variables} into \eqref{eq:dim-sys} and using 
\begin{subequations}\label{eq:non-dim-sys-normalized}
\begin{equation}\label{eq:non-dim-sys-normalized-N}
		\frac{dN(t)}{dt} = (b(N(t))  - m -  wI(t))N(t),
\end{equation}
we obtain the non-dimensionalized eco-evolutionary SIR model
	\begin{align}
		&\frac{dS(t)}{dt} = b(N(t))  -   (b(N(t))  +\Ro I(t)	-  wI(t))S(t), \label{eq:non-dim-sys-normalized-S}\\
		& \frac{\partial i(x,t)}{\partial t} = \ep^2 \frac{\partial^2i(x,t)}{\partial x^2} +G_1(x,t) i(x,t), \label{eq:non-dim-sys-normalized-i}\\
		& \frac{\partial r(x,t)}{\partial t} = -G_2(x,t)r(x,t) + (1-m-w)i(x,t), \label{eq:non-dim-sys-normalized-r}
	\end{align}
\end{subequations}
where
\begin{subequations}
\begin{gather}
	b(N) = b_0\left(1-\tfrac{N}{Q}\right),\qquad K(x) = \dimvar{K}\left(\tfrac{x}{\dimvar{a}}\right), \label{eq:non-dim-birth/kernel}\\
    b_0 := \frac{\dimvar{b}_0}{\dimvar{\tau}},\qquad \Ro = \frac{\dimvar{\beta}_0}{\dimvar{\tau}},\qquad m = \frac{\dimvar{\mu}}{\dimvar{\tau}},\qquad w = \frac{\dimvar{\omega}}{\dimvar{\tau}},\qquad \ep^2  = \frac{\dimvar{a}^2 \dimvar{D}}{\dimvar{\tau}}. \label{eq:non-dim-parameters}
\end{gather}
\end{subequations}
and where the functions
\begin{subequations}\label{eq:G1-G2-def}
	\begin{align}
		& G_1(x,t) = \Ro\left(\int_{-\infty}^x K(x-y)r(y,t)dy + S(t)\right) + m + wI(t) - b(N(t)) - 1, \label{eq:G1-def}\\
		& G_2(x,t) = \Ro\int_{x}^\infty K(y-x)i(y,t)dy + b(N(t)) - wI(t),\label{eq:G2-def}
	\end{align}
\end{subequations}
nonlinearly couple $S(t)$, $i(x,t)$, and $r(x,t)$. 

The dimensionless parameters $b_0$, $\Ro$, $m$, and $w$ above respectively describe the ratio of the average infection duration to the average time between births, contacts, natural deaths, and infection-induced deaths. In particular larger values of $m$ reflect longer lasting infections, while larger values of $w$ reflect more deadly infections. These two parameters are constrained by $0<m+w<1$ and, since $m+w=(\mu+\omega)/(\mu+\nu+\omega)$, the limit $m+w\rightarrow 1^-$ corresponds to lifelong infections (i.e.\@ the recovery rate $\nu$ vanishes). Finally, the dimensionless parameter $\ep>0$ can be interpreted as the strength of immune evasion throughout an average infection. To see this, denote by $\dimvar{\Delta x_\mathrm{mut}^2}$ and $\dimvar{\Delta t_{\mathrm{mut}}}$ the average mean-squared displacement in antigenic space and average time between mutations respectively. Also let $\dimvar{\tau}^{-1} = \dimvar{\Delta t_\mathrm{inf}}$ be the average infection duration and let $\dimvar{a} = \dimvar{\Delta K}/\dimvar{\Delta x} $. Then
\begin{equation*}
    \ep^2 = \frac{\dimvar{\Delta x_\mathrm{mut}^2}}{\dimvar{\Delta t_\mathrm{mut}}}\left(\frac{\dimvar{\Delta K}}{\dimvar{\Delta x}}\right)^2\dimvar{\Delta t_\mathrm{inf}}.
\end{equation*}
The ratio $\frac{\dimvar{\Delta t_\mathrm{inf}}}{\dimvar{\Delta t_\mathrm{mut}}}$ counts the average number of mutations per infection, while $\frac{\dimvar{(\Delta K)^2}\dimvar{\Delta x_\mathrm{mut}^2}}{(\dimvar{\Delta x})^2}$ is a measure of the mean-square increase in immune evasion per mutation. 

We conclude by noting that $\Ro$ is the basic reproduction number and is assumed to satisfy $\Ro>1$. Indeed, observe that linearizing \eqref{eq:non-dim-sys-normalized} about the trivial steady state $(S,i,r)=(1,0,0)$ yields the equation
\begin{equation*}
    \frac{\partial \varphi}{\partial t} = \ep^2\frac{\partial^2\varphi}{\partial x^2} + (\Ro - 1)\varphi,
\end{equation*}
which reflects the initial \textit{exponential growth} after a perturbatively small increase in the infected class, provided that $\Ro>1$.

\section{Traveling Pulse Solutions}\label{sec:traveling-pulse}

\begin{figure}[t!]
    \centering
    \includegraphics[width=0.95\linewidth]{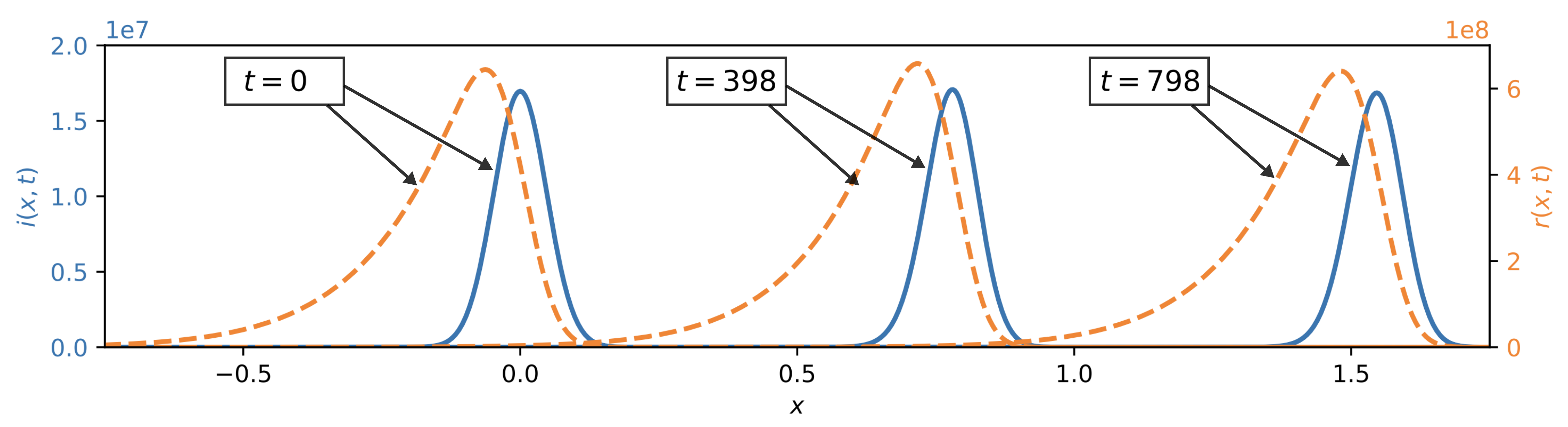}
    \caption{Numerical simulation of \eqref{eq:non-dim-sys-normalized} for parameters $Q=3\times10^8$, $b_0=2$, $\Ro=2$, $\ep=10^{-1}$, $m=1\times10^{-2}$, and $w=1\times10^{-6}$. The solid blue (resp. orange dashed) curves indicate the infected (resp. recovered) population distributions over antigenic space at the indicated times.}
    \label{fig:intro-sample-sim}
\end{figure}

Numerical simulations of \eqref{eq:non-dim-sys-normalized} suggest that when $\mathcal{R}_0>1$, small perturbations destabilize the spatially homogeneous steady state $S=1$ and $i,r=0$. Moreover, after an initial transient the solution appears to settle to an endemic equilibrium in which $S$ is constant while $i(x,t)$ and $r(x,t)$ are traveling pulses with the latter lagging behind the former (see Figure \ref{fig:intro-sample-sim}). In this section we provide an asymptotic analysis of these traveling pulse solutions valid for a range of $m$ and $w$ values extending that previously considered by Lin et.\@ al.\@ in \cite{lin_2003}.

Letting $c>0$ be the speed of the traveling pulses we seek a solution of the form
\begin{equation*}
    S = S_e,\qquad i(x,t) = i_e(x-ct),\qquad r(x,t) = r_e(x-ct).
\end{equation*}
Note that the total proportions of infected and recovered individuals are respectively given by
\begin{equation*}
    I_e = \int_{-\infty}^\infty i_e(u)du,\qquad R_e = \int_{-\infty}^\infty r_e(u)du,
\end{equation*}
both of which are time-independent. Thus, since each of $S_e$, $I_e$, and $R_e$ are time-independent, so too is the total population size $N=N_e$. We solve \eqref{eq:non-dim-sys-normalized-N} and \eqref{eq:non-dim-sys-normalized-S} explicitly as
\begin{equation}\label{eq:endemic-N-S}
    N_e = \left(1 - \frac{m+wI_e}{b_0}\right)Q,\qquad S_e = \frac{m+wI_e}{m+\Ro I_e}.
\end{equation}
Moreover, in the moving reference frame $u=x-ct$ equations \eqref{eq:non-dim-sys-normalized-i} and \eqref{eq:non-dim-sys-normalized-r} become
\begin{subequations}\label{eq:i_e_r_e_eq}
\begin{align}
	&\ep^2 i_e''(u) + c i_e'(u) + g_1(u)i_e(u) = 0, & -\infty<u<\infty, \label{eq:i_e_eq}\\
	& cr_e'(u) - g_2(u)r_e(u) + (1-m-w)i_e(u) = 0, & -\infty<u<\infty, \label{eq:r_e_eq}
\end{align}
\end{subequations}
where
\begin{subequations}\label{eq:g_1_g_2_def}
\begin{align}
	& g_1(u) := \Ro\left(\int_{-\infty}^u K(u-v)r_e(v)dv + S_e\right) - 1, \label{eq:g_1_def}\\
    & g_2(u) := \Ro\int_{u}^\infty K(v-u)i_e(v)dv + m. \label{eq:g_2_def}
\end{align}
\end{subequations}

We assume that $i_e(u)$ and $r_e(u)$ are positive for all $u\in\mathbb{R}$ and that $i_e,r_e\rightarrow 0$ as $u\rightarrow\pm\infty$. These conditions are used to deduce the speed $c>0$ as discussed below. Under these assumptions on $i_e(u)$ and $r_e(u)$ we integrate \eqref{eq:i_e_eq} and \eqref{eq:r_e_eq} over $\mathbb{R}$ and use $0\leq K(z)\leq 1$ to obtain the bounds
\begin{equation}\label{eq:tot-prop-bounds}
    0\leq S_e\leq \frac{1}{\Ro},\quad 0\leq I_e\leq 1- \frac{1}{\Ro},\quad \frac{1-m-w}{m + \Ro I_e}I_e\leq R_e\leq \frac{1-m-w}{m}I_e.
\end{equation}
We also find that
\begin{subequations}\label{eq:g1_prop}
\begin{equation}\label{eq:g1'_and_g1_lim}
    g_1'(u) = \Ro\int_{-\infty}^{u}K'(u-v)r_e(v)dv>0,\quad
	g_1(u) \rightarrow \begin{cases} \Ro S_e - 1 \leq 0, & u\rightarrow-\infty, \\ \Ro(1-I_e) - 1 \geq 0, & u\rightarrow+\infty.\end{cases}
\end{equation}
Therefore $g_1(u)$ is monotone increasing and changes sign exactly once due to the inequalities in \eqref{eq:tot-prop-bounds}. Using the translation invariance of \eqref{eq:i_e_r_e_eq} we \textit{impose}, without loss of generality,  that
\begin{equation}
    g_1(0) = 0.
\end{equation}
\end{subequations}

\begin{figure}
    \centering
    \includegraphics[width=0.9\linewidth]{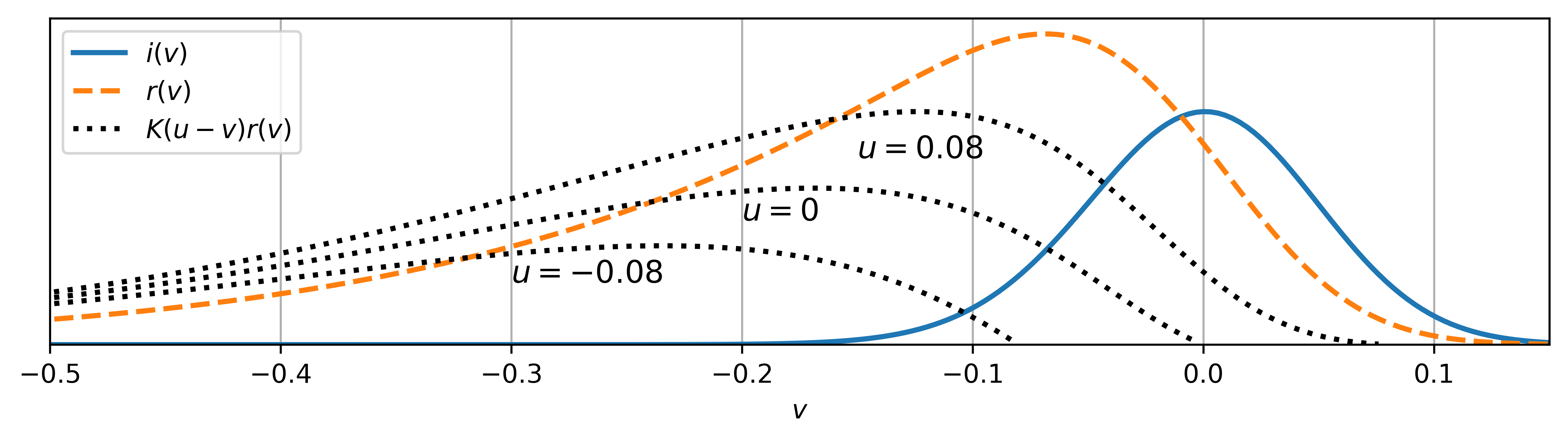}
    \caption{Schematic of the endemic infected (solid blue curve) and recovered (dashed orange curve) distributions in a moving reference frame. Superimposed are the curves $K(u-v)r(v)$ (dotted black curves) at indicated values of $u$, whose integral determines the infected growth rate due to reinfection. The quantities $i_e(v)$, $r_e(v)$, and $K(u-v)r_e(v)$ have been rescaled for illustrative purposes. The rescaling for $K(u-v)r_e(v)$ is constant for each choice of $u$ to highlight their relative magnitudes.}
    \label{fig:diagram-r-infectivity-all.pdf}
\end{figure}

The function $g_1(u)$ describes the growth rate of the infected population in the moving reference frame. The quantitative properties of $g_1(u)$ depend nonlinearly on the solution to \eqref{eq:i_e_r_e_eq}, but \eqref{eq:g1_prop} alone determines its qualitative behavior.  Specifically, from \eqref{eq:g1'_and_g1_lim} we see that growth of the infected population due to infection of the susceptible population as $u\rightarrow-\infty$ (i.e.\@ for strains far in the past) is insufficient to counteract its decay due to recovery and death. The change in sign of $g_1(u)$ across $u=0$ is therefore due to reinfection of the recovered population, as is most transparent upon using $S_e+I_e+R_e=1$ to write
\begin{equation*}
    \lim_{u\rightarrow+\infty} g_1(u) = \Ro R_e + \Ro S_e - 1 \geq 0.
\end{equation*}
It is therefore antigenic drift and reinfection of the recovered class that are responsible for pulse propagation in the eco-evolutionary SIR model \eqref{eq:dim-sys}. This coupling is the mechanism described in \S\ref{sec:intro}, stated here in terms of the model: the recovered density $r_e$ enters $g_1(u)$ through the convolution in \eqref{eq:g_1_def}, so the same quantity that lowers $g_1(u)$ at $u<0$, where hosts are immune, raises it at $u>0$, where they are only partially so. Reinfection of the recovered class maintains the endemic state, through the term $\Ro\int_{-\infty}^u K(u-v)r_e(v)dv$, rather than the entry of newborn hosts into the susceptible class. In Figure \ref{fig:diagram-r-infectivity-all.pdf} we plot $i_e(v)$, $r_e(v)$, and $K(u-v)r_e(v)$ (rescaled for readability) which shows that an increase in $u$ expands the recovered population that is susceptible to reinfection by strain $u$. Consequently the growth rate due to reinfection, $\int_{-\infty}^u K(u-v)r_e(v)dv$, increases monotonically. Therefore, when coupled with antigenic drift (i.e.\@ diffusion in antigenic space), reinfection of the recovered class biases growth to greater values of $x$ in antigenic space, so the pulse advances. Finally, we comment that mathematically the change in sign of $g_1(u)$ is crucial for the existence of a homoclinic solution to \eqref{eq:i_e_r_e_eq}.

While the change in sign of $g_1(u)$ provides a mechanistic description for pulse propagation, the rate at which $g_1(u)$ changes sign, i.e.\@ $g_1'(0)$, determines the spread of the infected distribution in antigenic space. Specifically, if $g_1'(0)$ is small then the transition from decay to growth at $u=0$ is small, leading to a wider infected distribution, with the opposite being true if $g_1'(0)$ is large. Properties of the cross-immunity kernel together with \eqref{eq:g1'_and_g1_lim} yield the bounds
\begin{equation*}
    0 < g_1'(0) \leq \Ro R_e.
\end{equation*}
With $\Ro$ fixed, the lower bound is approached as $R_e\rightarrow 0$. Alternatively, since $K'(-u)\rightarrow 0$ as $u\rightarrow-\infty$ the lower bound can also be obtained if there is a sufficiently large lag between the dominant infected and recovered strains. On the other hand, since $K'(0)=1$ the upper bound is approached as the lag between the dominant infected and recovered strains is diminished, provided that all of $r_e(u)$ is concentrated at $u<0$.

The observations above follow from \eqref{eq:g1_prop} alone, and yield only qualitative properties of solutions to the eco-evolutionary SIR model \eqref{eq:dim-sys}. In the following sections we analyze solutions to \eqref{eq:i_e_r_e_eq} and systematically corroborate the above heuristics. First, in \S\ref{subsec:traveling-pulse-speed} we obtain an expression for the speed $c_0$ (previously obtained by Lin et al.\@ in \cite{lin_2003}) using a standard phase-plane type argument. Next, for $\ep\ll 1$ we use the WKB method in \S\ref{subsec:asymptotics-of-traveling-sol} to fully characterize $i_e(u)$ and $r_e(u)$ in terms of the two \textit{scalars} $I_\asy$ and $\sigma_\asy^2$ which respectively correspond to the total infected population and its variance. This asymptotic analysis leads to a nonlinear system of two algebraic equations in the unknowns $I_\asy$ and $\sigma_\asy^2$. Moreover, the inverse proportionality between $g_1'(0)$ and $\sigma_\asy^2$ is explicitly borne out of the asymptotic analysis in \S\ref{subsec:asymptotics-of-traveling-sol}. In \S\ref{subsec:identities} we derive directly from \eqref{eq:non-dim-sys-normalized} counterparts to the Price equation and Fisher's fundamental theorem of natural selection, and show that leading order forms of these expressions correspond to the nonlinear system derived in \S\ref{subsec:asymptotics-of-traveling-sol}. In \S\ref{subsec:m-regimes} we then consider the $m\gg\sqrt{\ep}$ and $\ep^2\ll m\ll \ep$ regimes which lead to simplified expressions for $I_\asy$ and $\sigma_\asy^2$. Finally, in \S\ref{subsec:exponential-kernel} we introduce the exponential cross-immunity kernel which we use as an example for numerical calculations in subsequent sections. 


\subsection{Traveling Pulse Speed}\label{subsec:traveling-pulse-speed}

A lower bound on the traveling pulse's speed is determined by considering the limiting behavior of \eqref{eq:i_e_eq} as $u\rightarrow\pm\infty$ and requiring that $i_e(u)$ is strictly positive and decaying to zero. Letting $u\rightarrow\pm\infty$ in \eqref{eq:i_e_eq} and using \eqref{eq:g1_prop} we find that $i_e(u)\sim C_\pm e^{\gamma_\pm u}$ as $u\rightarrow\pm \infty$ where $\gamma_+$ and $\gamma_-$ respectively satisfy the quadratics
\begin{equation*}
    \ep^2\gamma_-^2 + c\gamma_- + \Ro S_e - 1 = 0,\quad \ep^2\gamma_+^2 + c\gamma_+ + \Ro(1-I_e) - 1 = 0.
\end{equation*}
Since $\Ro S_e  - 1\leq 0$ the first quadratic always has a positive real-valued solution $\gamma_-$. On the other hand, since $\Ro(1-I_e)-1\geq 0$, for $\gamma_+$ to be negative and real valued we require that
\begin{equation}\label{eq:c_lower_bound}
	c \geq \ep c_0,\qquad\text{where}\quad c_0:= 2\sqrt{\Ro (1-I_e) - 1}.
\end{equation}
We assume that marginal stability criteria holds so that the minimum speed is selected, i.e.\@ $c=\ep c_0$. In addition, we assume that $c_0=O(1)$, which in particular means that $I_e$ is sufficiently far from its upper bound in \eqref{eq:tot-prop-bounds}. We note two features of \eqref{eq:c_lower_bound}. First, $c_0$ is determined by $\gamma_+$, and hence by $i_e(u)$ as $u\rightarrow+\infty$, where $i_e$ is exponentially small and $g_1(u)\rightarrow\Ro(1-I_e)-1>0$. The speed is therefore fixed by the tail of the infected distribution at large $u$ and not by its bulk near $u=0$. Second, $c_0$ is decreasing in $I_e$. Writing $\Ro(1-I_e)-1 = \Ro(S_e+R_e)-1$ using $S_e+I_e+R_e=1$, the quantity that determines $c_0$ is the proportion of hosts not currently infected, since by assumption i.) on $\dimvar{K}$ these are the hosts that strains at large $u$ can infect. Thus $c$ and $I_e$ are not independent: both are unknowns of the same system \eqref{eq:i_e_r_e_eq}, and we determine them jointly in \S\ref{sec:results}. The joint determination of $c$ and $I_e$ is what we meant in \S\ref{sec:intro} by calling the model eco-evolutionary.

\subsection{Asymptotic Approximation of Traveling Solutions}\label{subsec:asymptotics-of-traveling-sol}

Our goal is to construct an asymptotic approximation of the traveling pulse solutions when $\ep\ll 1$. The system \eqref{eq:i_e_r_e_eq} constitutes a non-linear and nonlocal system in $i_e(u)$ and $r_e(u)$. However, we can leverage the smallness of $\ep$ to obtain accurate approximations. Note that independently of the specific form of $i_e(u)$ and $r_e(u)$, the function $g_1(u)$ must satisfy the properties \eqref{eq:g1_prop}. We therefore neglect the nonlinear and nonlocal dependencies of $g_1(u)$ on the unknown functions $i_e(u)$ and $r_e(u)$, and instead treat it as a known function of $u$ satisfying \eqref{eq:g1_prop}. This renders \eqref{eq:i_e_eq} a \textit{linear} equation in $i_e(u)$, and in particular one well suited for a classical WKB-type analysis. Letting $i_e(u) = \exp\left(-\frac{c_0}{2\ep}u\right) \varphi(u)$ and substituting into \eqref{eq:non-dim-sys-normalized-i} gives 
\begin{equation*}
	\ep^2 \varphi''(u) = Q(u)\varphi(u),\quad -\infty<u<\infty,\qquad\text{where}\quad Q(u) := \frac{c_0^2}{4} - g_1(u).
\end{equation*}
Since $Q(u)$ is a monotone decreasing function with $Q(u)\rightarrow 0$ as $u\rightarrow \infty$ this problem is well suited for a WKB asymptotic approximation provided $Q(u)\geq O(1)$. We assume this is the case for an interval of the form $-\infty<u\leq L$ for some positive $L=O(1)$. Thus we obtain the asymptotic approximation
\begin{equation*}
    i_e(u) \sim \frac{C_\ep}{[Q(u)]^{1/4}}\exp\left(-\frac{c_0}{2\ep}u-\frac{1}{\ep}\int_u^L\sqrt{Q(v)}dv\right),
\end{equation*}
where $C_\ep>0$ is some normalization constant.

Anticipating that $i_e(u)$ is concentrated around $u=0$, we calculate
\begin{equation*}
	\int_u^L \sqrt{Q(v)} dv = \int_0^L \sqrt{Q(v)}dv - \sqrt{Q(0)} u - \tfrac{Q'(0)}{4\sqrt{Q(0)}}u^2 + O(u^3),\qquad \text{for}\quad |u|\ll1
\end{equation*}
and note that $Q(0)=c_0^2/4$ and $Q'(0)=-g_1'(0)$. We thus obtain $i_e(u)\sim i_\asy(u)$ where
\begin{subequations}
\begin{equation}\label{eq:i_e_gaussian}
	i_{\asy}(u) = \frac{I_\asy}{\sigma_\asy\sqrt{2\pi}}\exp\left(-\frac{u^2}{2\sigma_\asy^2}  \right),
\end{equation}
and where we define
\begin{equation}\label{eq:i_e_variance_speed}
    \sigma_\asy^2 := \frac{c_\asy}{g_1'(0)}\quad\text{and}\quad c_\asy := 2\ep\sqrt{\Ro(1-I_\asy)-1}.
\end{equation}
\end{subequations}
The proportion of infected individuals is therefore approximately normally distributed and completely characterized by the two parameters $I_\asy$ and $\sigma_\asy^2$ corresponding, respectively, to the total proportion of infected individuals and its variance in antigenic space. Moreover, \eqref{eq:i_e_variance_speed} systematically justifies the heuristic argument that the variance is inversely proportional to $g_1'(0)$ in \S\ref{sec:traveling-pulse} based on \eqref{eq:g1'_and_g1_lim}.

The expression for $i_\asy(u)$ is based on the assumption that $|u|\ll 1$ and is therefore valid provided that $\sigma_\asy\ll 1$. We impose the stronger assumption that in fact $g_1'(0)=O(1)$ so that $\sigma_\asy=O(\sqrt{\ep})$. Note that this assumption is a constraint on the distribution of recovered individuals. Together with the assumption $c_\asy=O(\ep)$, this ultimately leads us to restrict the parameter regime over which our asymptotic approximation is valid. In order to determine the unknowns $I_\asy$ and $\sigma_\asy$ we turn next to approximating the recovered distribution.

First, observe that with $i_e(u)\sim i_\asy(u)$ we have $g_2(u) \sim \Ro I_\asy K(-u) + m$. Substituting into \eqref{eq:r_e_eq} we determine that the asymptotic approximation $r_e(u)\sim r_\asy(u)$ satisfies
\begin{subequations}\label{eq:r_asy_eq_ode}
\begin{align}
	&  r_\text{asy}'(u) - \frac{\Ro I_\text{asy} K(-u) + m}{c_\asy} r_\text{asy}(u)  + \frac{1-m-w}{c_\asy}i_\text{asy}(u) = 0, & -\infty<u<\infty, \label{eq:r_asy_eq_ode_0}\\
    & r_\asy(u)\rightarrow 0, & u\rightarrow+\infty, \label{eq:r_asy_eq_ode_1}
\end{align}
which we can solve explicitly as
\begin{equation}\label{eq:r_asy_exact}
    r_\asy(u) = \frac{1-m-w}{c_\asy\sigma_\asy\sqrt{2\pi}}I_\asy\int_u^\infty \exp\left(-\tfrac{v^2}{2\sigma_\asy^2} - \tfrac{m}{c_\asy}(v-u) - \tfrac{\Ro I_\asy}{c_\asy}\int_u^v K(-z)dz \right)dv.
\end{equation}
\end{subequations}

It remains only to determine the two unknown parameters $I_\asy$ and $\sigma_\asy$ in terms of which both $i_\asy(u)$ and $r_\asy(u)$ are parametrized. We obtain two equations for $I_\asy$ and $\sigma_\asy$ by imposing that $g_1(0)=0$ and $g_1'(0)=c_\asy/\sigma_\asy^2$ which respectively give
\begin{subequations}\label{eq:nonlin_sys_0}
	\begin{align}
		& \int_0^{\infty} K(u)r_\asy(-u)du + \tfrac{m + w I_\asy}{m + \Ro I_\asy} - \tfrac{1}{\Ro} = 0, \label{eq:nonlin_sys_0_1}\\
		& \sigma_\asy^2\Ro \int_0^{\infty} K'(u)r_\asy(-u)du - c_\asy  = 0. \label{eq:nonlin_sys_0_2}
	\end{align}
\end{subequations}

Solving the system \eqref{eq:nonlin_sys_0} for the unknowns $I_\asy$ and $\sigma_\asy$ yields asymptotic approximations for the infected and recovered distributions via \eqref{eq:i_e_gaussian} and \eqref{eq:r_asy_exact} respectively. While the asymptotic reduction leading to \eqref{eq:nonlin_sys_0} provides insights and simplifications, the system \eqref{eq:nonlin_sys_0} must in general be solved numerically. Furthermore, the numerical solution of \eqref{eq:nonlin_sys_0} is not without its own challenges, stemming primarily from the unknown $\ep$-dependence of $I_\asy$ and resulting potential for multiple spatial scales in the exponent appearing in \eqref{eq:r_asy_exact}. In \S\ref{subsec:m-regimes} we address this difficulty by considering two distinguished parameter regimes for the parameter $m$ under which \eqref{eq:nonlin_sys_0} is further simplified. In \S\ref{subsec:exponential-kernel} we introduce the special case of an exponential cross-immunity kernel for which the integrals in \eqref{eq:nonlin_sys_0} can be quickly evaluated. First, however we conclude with a few remarks on the above asymptotic analysis.

\begin{rem}\label{rem:mass-conservation}
	The asymptotic solution constructed above respects mass conservation. To see this let $I_\asy$ and $\sigma_\asy$ solve \eqref{eq:nonlin_sys_0_1}. Integrating \eqref{eq:r_asy_eq_ode_0} over $-\infty<u<\infty$ and using \eqref{eq:nonlin_sys_0_1} for $\int_{-\infty}^0K(-u)r_\asy(u)du$ then yields {\small$R_\asy = \tfrac{I_\asy}{m}\left(\tfrac{m+wI_\asy}{m+\Ro I_\asy} \Ro  - m - w\right)$}. With \eqref{eq:endemic-N-S} for $S_\asy$ it is then straightforward to show that $S_\asy+I_\asy+R_\asy=1$.
\end{rem}

\begin{rem}\label{rem:ode-for-rho}
    Since $K(z)=0$ for all $z\leq 0$ we can solve \eqref{eq:r_asy_eq_ode} explicitly for $u\geq 0 $ to get
    \begin{equation}\label{eq:r_asy_pos}
        r_\text{asy}(u) = \tfrac{1-m-w}{2 c_\asy}I_\text{asy}  e^{\frac{m}{c_\asy}u + \frac{m^2\sigma_\asy^2}{2c_\asy^2}}\erfc\left(\tfrac{u}{\sigma_\asy\sqrt{2}} + \tfrac{\sigma_\asy m}{c_\asy\sqrt{2}}   \right),\qquad u\geq 0.
    \end{equation}
    The remaining values of $r_\asy(u)$ for $u<0$ are then found by solving an initial-value-problem. Note that {\small$r_\text{asy}(0) = \tfrac{1-m-w}{2 \sqrt{\pi}c_\asy}I_\text{asy}  U\left(\tfrac{m^2\sigma^2}{2c_\asy^2} \right)$} where $U\left(z\right) := \sqrt{\pi}e^{z}\erfc(\sqrt{z})$. Thus
    \begin{subequations}\label{eq:r_asy_neg}
    \begin{equation}\label{eq:r_asy_rescale_0}
        r_\asy(u) = \frac{1-m-w}{2c_\asy\sqrt{\pi}}I_\asy U\left(\tfrac{m^2\sigma_\asy^2}{2c_\asy^2}\right) \rho(u),
    \end{equation}
    where $\rho(u)$ solves
    \begin{align}
        & \rho'(u) - \left(\tfrac{\Ro I_\asy}{c_\asy}K(-u) + \tfrac{m}{c_\asy}\right)\rho(u) = -\tfrac{\sqrt{2}}{\sigma_\asy}\tfrac{\exp\left(-\tfrac{u^2}{2\sigma_\asy^2}\right)}{U\left(\tfrac{m^2\sigma_\asy^2}{2c_\asy^2}\right)}, & u< 0\label{eq:r_asy_rescale_1a}\\
        & \rho(0) = 1. \label{eq:r_asy_rescale_1b}
    \end{align}
    \end{subequations}
\end{rem}

\subsection{Exact Identities for the Speed and the Variance in Growth Rate}\label{subsec:identities}

The asymptotic construction of \S\ref{subsec:asymptotics-of-traveling-sol} closes on the two conditions $g_1(0)=0$ and $g_1'(0)\sigma_\asy^2=c_\asy$. The condition $g_1(0)=0$ is a normalization. The growth rate $g_1(u)$ changes sign exactly once by \eqref{eq:g1_prop}, and the translation invariance of \eqref{eq:i_e_r_e_eq} allows us to place that sign change at $u=0$, which fixes the origin of the moving coordinate and imposes no constraint on the solution. The condition $g_1'(0)\sigma_\asy^2=c_\asy$, which relates the speed $c_\asy$ of antigenic advance to the antigenic variance $\sigma_\asy^2$ and to the slope $g_1'(0)$ of the growth rate at the sign change, is instead the leading-order form of an identity that holds exactly for the traveling pulse. That identity and two others obtained in the same way are the counterparts in our model of the Price equation \cite{price_1970,price_1972} and of Fisher's fundamental theorem of natural selection \cite{fisher_1930}.

Write $\langle f\rangle := I_e^{-1}\int_{-\infty}^\infty f(u)i_e(u)du$ for the average of a quantity $f$ over the infected population. Multiplying \eqref{eq:i_e_eq} by $1$, by $u$, and by $g_1(u)$, and integrating over $-\infty<u<\infty$, gives respectively 
\begin{subequations}\label{eq:moments}
\begin{align}
    & \langle g_1\rangle = 0, \label{eq:moment-mean}\\
    & \mathrm{cov}\!\left[g_1,u\right] = c, \label{eq:moment-cov}\\
    & \mathrm{var}\!\left[g_1\right] + \ep^2\langle g_1''\rangle
        = c\,\langle g_1'\rangle. \label{eq:moment-var}
\end{align}
\end{subequations}
The identities \eqref{eq:moments} follow from \eqref{eq:i_e_eq} together with the decay of $i_e(u)$ as $u\rightarrow\pm\infty$, and so hold with no approximation in $\ep$.

The quantity $g_1(u)$, the per-capita growth rate of the variant at antigenic position $u$ in the moving frame, is the fitness of that variant, and the averages in \eqref{eq:moments} are taken over the circulating variants weighted by their prevalence. The first identity \eqref{eq:moment-mean} states that the average growth rate across those variants vanishes, which is the stationarity of the total infected proportion $I_e$. Since $g_1(u)$ changes sign once, at $u=0$, the variants ahead of the sign change are growing and those behind it are
declining, in exact compensation.

The second identity \eqref{eq:moment-cov} ties the speed of antigenic evolution to the spread of the infected population in antigenic space. The Price equation, a very general identity of population genetics that Day and Gandon \cite{day_2007} have applied to epidemiological models, states that the rate of change in the mean of a trait equals the covariance between that trait and fitness, together with a term accounting for any bias in the transmission of the trait between generations. Taking antigenic position $u$ as the trait and the growth rate $g_1(u)$ as the fitness, the transmission term vanishes here because antigenic mutation is modeled as unbiased diffusion, and what remains is \eqref{eq:moment-cov}: the speed at which the pathogen advances through antigenic space equals the covariance between a variant's growth rate and its antigenic position. The speed is therefore set jointly by how far the circulating variants spread in antigenic space and by how steeply the growth rate increases with position, and not by the rate of antigenic mutation alone.

Approximating $i_e(u)$ by the Gaussian \eqref{eq:i_e_gaussian} and $g_1(u)$ by $g_1'(0)u$ near $u=0$ reduces the covariance in \eqref{eq:moment-cov} to $g_1'(0)\sigma_\asy^2$ and returns $c_\asy=g_1'(0)\sigma_\asy^2$. The closure condition \eqref{eq:i_e_variance_speed} is therefore the leading-order form of \eqref{eq:moment-cov}. 

The third identity \eqref{eq:moment-var} is the counterpart of Fisher's fundamental theorem of natural selection, which states that the rate of increase in a population's mean fitness attributable to selection equals the variance in fitness among its members. For the traveling pulse $G_1(x,t)=g_1(x-ct)$, so that $\partial_tG_1=-c\,g_1'(x-ct)$, and \eqref{eq:moment-var} becomes
\begin{equation}\label{eq:fisher-balance}
    \mathrm{var}\!\left[g_1\right]
    + \ep^2\big\langle \partial_x^2 G_1\big\rangle
    = -\big\langle \partial_t G_1\big\rangle.
\end{equation}
The variance in growth rate on the left of \eqref{eq:fisher-balance} is the quantity appearing in Fisher's theorem, and it is the rate at which selection raises the average growth rate. The second term on the left is the contribution of antigenic mutation, of order $\ep^2$, which does not enter at leading order. The right-hand side is the rate at which the growth rate at each fixed antigenic position falls as the recovered density at that position accumulates. At leading order \eqref{eq:moment-var} is \eqref{eq:moment-cov} multiplied by $g_1'(0)$, so the two identities express one condition, in units of antigenic position and in units of growth rate respectively.

Fisher observed that the total change in a population's mean fitness carries, besides the variance in fitness, a further term accounting for change in the environment, which he called its deterioration. The identity \eqref{eq:fisher-balance} is that decomposition for our model, with two features particular to it. The term $-\langle\partial_tG_1\rangle$ is not imposed exogenously but produced by the pathogen, since the recovered distribution $r_e$ that determines the growth rate $g_1(u)$ through \eqref{eq:g_1_def} is produced by the infections the pathogen has already caused. And the two sides of \eqref{eq:fisher-balance} balance exactly, so that the average growth rate stays at zero: selection raises the average growth rate at precisely the rate at which the immunity accumulated at the positions already visited lowers it. The endemic traveling pulse is the state in which those two rates coincide.

\subsection{Asymptotics in the $m\gg\sqrt{\ep}$ and $\ep^2\ll m\ll \ep$ Regimes}\label{subsec:m-regimes}

Additional analysis of \eqref{eq:r_asy_neg} reveals that $m\gg\sqrt{\ep}$ and $\ep^2\ll m\ll \ep$ are distinguished asymptotic regimes. Consequently we obtain further asymptotic approximations for the recovered distribution as well as for the unknowns $I_\asy$ and $\sigma_\asy$ which we summarize below.

In the $m\gg \sqrt{\ep}$ regime, corresponding to long-lasting infections, we find that $r_\asy(u)$ is approximately normally distributed and given by \eqref{eq:r_asy_pos_asy}. Substituting into \eqref{eq:nonlin_sys_0} then gives leads to the following leading order asymptotics (see Appendix \ref{app:asymptotics-large-m}) 
\begin{equation}\label{eq:large-m-results}
    I_\asy \sim \tfrac{m}{1-w}\left(1-\frac{1}{\Ro}\right),\quad \sigma_\asy^2 \sim 4\ep \sqrt{\tfrac{1-w}{(\Ro - 1)(1-m-w)}},\quad c_\asy \sim 2\ep\sqrt{\tfrac{(\Ro-1)(1-m-w)}{1-w}}.
\end{equation}
Equation \eqref{eq:large-m-results} supports two distinct comparisons. Since $c_\asy\sigma_\asy^2 = 8\ep^2$, fixing $\ep$ confines the pair $(c_\asy,\sigma_\asy^2)$ to a single hyperbola, so that any change in $m$, $w$, or $\Ro$ that lowers $c_\asy$ raises $\sigma_\asy^2$ by the reciprocal factor. On the other hand, varying $\ep$ moves between hyperbolae, and since $c_\asy=O(\ep)$ and $\sigma_\asy^2=O(\ep)$, both quantities then change in the same direction. The relationship between $c_\asy$ and $\sigma_\asy^2$ therefore depends on which parameter distinguishes the two cases compared. We return to both comparisons in \S\ref{sec:discussion}.

In the  $\ep^2\ll m\ll \ep$ regime, corresponding to short-lasting infections, we find that $r_\asy(u)$ can be approximated by the composite solution \eqref{eq:m-small-asy-r-comp}. Defining
\begin{equation}\label{eq:script-K-H-def}
    \mathcal{K}(\zeta) := \int_0^\infty e^{-\zeta\int_0^uK(z)dz}du,\quad \mathcal{H}(\zeta) := \int_0^\infty K'(u)e^{-\zeta\int_0^uK(z)dz}du,
\end{equation}
we then find the leading order asymptotics (see Appendix \ref{app:asymptotics-small-m})
\begin{subequations}
\begin{equation}\label{eq:small-m-asymptotics}
    I_\asy \sim 2\ep \zeta_\star \frac{\sqrt{\Ro -1}}{\Ro},\quad \sigma_\asy^2 \sim \frac{2\ep\sqrt{\Ro -1}}{(1-w)\zeta_\star\mathcal{H}(\zeta_\star)},\quad c_\asy\sim 2\ep\sqrt{\Ro-1},
\end{equation}
where $\zeta_\star$ is the unique solution to
\begin{equation}\label{eq:small-m-nonlinear-eq}
    \zeta\mathcal{K}\left(\zeta\right) - \frac{\Ro - w}{1-w} =0.
\end{equation}
\end{subequations}
In addition we find that the separation between the peaks of the infected and recovered populations is approximately given by \eqref{eq:m-small-peak-sep} and in particular scales like $O(\sqrt{\ep|\log\ep|})$. Thus the lag between dominant strains in the infected and recovered population decreases as $m$ increases.

\subsection{An Exponential Cross-Immunity Kernel}\label{subsec:exponential-kernel}

The asymptotics obtained in the preceding sections suggest that $\sigma_\asy=O(\sqrt{\ep})$ for all parameter regimes, whereas $I_\asy=O(1)$ for $\sqrt{\ep}\ll m < 1$ and $I_\asy=O(\ep)$ for $m\ll \ep$. However, the scaling of $I_\asy$ in the intermediate regime $O(\ep)\leq m\leq O(\sqrt{\ep})$ remains a priori unknown, leading to numerical difficulties when evaluating the integrals appearing in \eqref{eq:nonlin_sys_0}. To facilitate the numerical solution fo the nonlinear system \eqref{eq:nonlin_sys_0} we introduce and restrict our attention to the following exponential cross-immunity kernel:
\begin{equation}\label{eq:exponential-kernel}
    K(z) = \begin{cases}  1-e^{-z}, & z>0, \\ 0, & z\leq 0,\end{cases}
\end{equation}
The choice of this kernel leads to three simplifications (see Appendix \ref{app:exponential-kernel} for details). First, it provides a fast and robust iterative method for calculating the integrals appearing in \eqref{eq:nonlin_sys_0}, replacing the nonlinear system with \eqref{eq:nonlin_sys_2}. Second it allows for the explicit calculation of the functions $\mathcal{K}(\cdot)$ and $\mathcal{H}(\cdot)$ in the $\ep^2\ll m \ll \ep$ regime in terms of the lower incomplete gamma function. Finally, it allows the original system \eqref{eq:non-dim-sys-normalized} to be \textit{localized} in the sense that the integrals appearing in \eqref{eq:non-dim-sys-normalized} can be replaced by two additional PDEs
\begin{subequations}\label{eq:G1-G2-pdes}
\begin{equation}\label{eq:G1-G2-pdes-eq}
    \frac{\partial^2 G_1(x,t)}{\partial x^2} + \frac{\partial G_1(x,t)}{\partial x}=\Ro r(x,t)\quad\text{and}\quad \frac{\partial^2 G_2(x,t)}{\partial x^2} - \frac{\partial G_2(x,t)}{\partial x}=\Ro i(x,t),
\end{equation}
together with the limiting values
\begin{equation}\label{eq:G1-lim}
	G_1(x,t)\rightarrow\begin{cases} \Ro S(t) + m + wI(t)-b(N(t))-1, & x\rightarrow -\infty, \\ \Ro(R(t)+S(t))+ m + wI(t)-b(N(t))-1, & x\rightarrow +\infty,\end{cases}
\end{equation}
and
\begin{equation}\label{eq:G2-lim}
    G_2(x,t)\rightarrow\begin{cases} \Ro I(t) + b(N(t)) - wI(t), & x\rightarrow -\infty, \\ b(N(t)) - wI(t), & x\rightarrow +\infty.\end{cases}
\end{equation}
\end{subequations}

\section{Results}\label{sec:results}

The nonlinear algebraic system \eqref{eq:nonlin_sys_0} derived in \S\ref{subsec:asymptotics-of-traveling-sol} yields a numerically efficient method for investigating the behavior of traveling pulse solutions to \eqref{eq:non-dim-sys-normalized} over a substantial portion of its parameter space. By restricting our attention to the exponential immunity kernel introduced in \S\ref{subsec:exponential-kernel}, the nonlinear algebraic system \eqref{eq:nonlin_sys_0} further reduces to \eqref{eq:nonlin_sys_2} for which each term is calculated in terms of the convergent series \eqref{eq:Q-series}. In this section we specifically focus on the exponential kernel and solutions to \eqref{eq:nonlin_sys_2}. We first explore the behavior of $I_\asy$ and $\sigma^2_\asy$ as $m$, $w$, and $\Ro$ are varied for fixed values of $\ep$. We further compare solutions to \eqref{eq:nonlin_sys_2} with the \textit{large}- and \textit{small}-$m$ asymptotics of \S\ref{subsec:m-regimes}, finding favorable agreement. Next we compare the profiles of the infected and recovered population densities predicted by the asymptotics, with those found by directly numerically simulating the system \eqref{eq:non-dim-sys-normalized}. We conclude by showing that under certain parameter regimes the \textit{total} infected population exhibits non-monotonic behavior with respect to infection-induced mortality. In particular, we illustrate that for sufficiently small values of the intrinsic birth rate, the total infected population is maximized for intermediate levels of infection-induced mortality.

\subsection{Solving the Nonlinear Algebraic System}\label{subsec:sol-nonlin}

\begin{figure}
    \centering
    \includegraphics[width=0.9\linewidth]{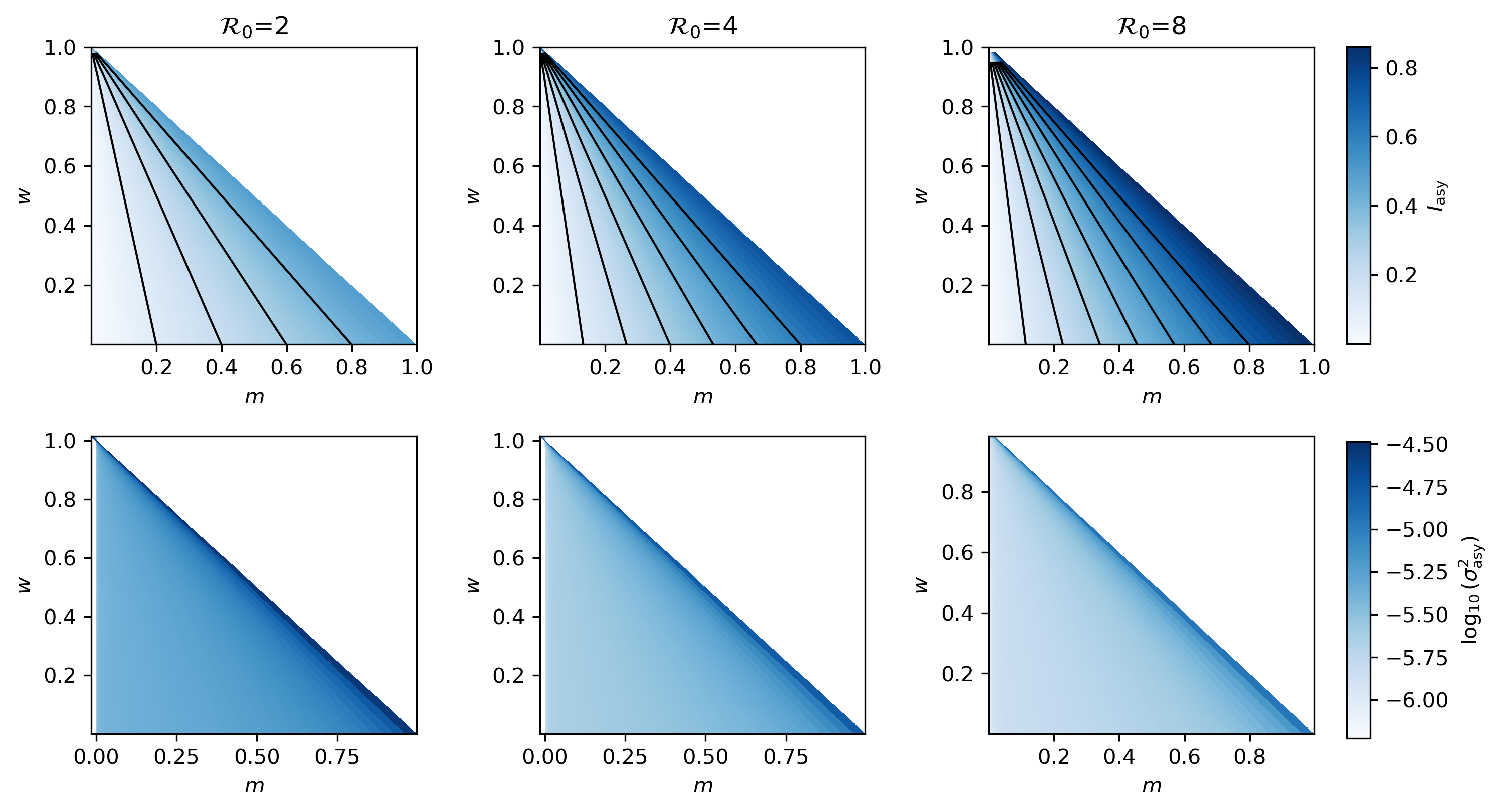}
    \caption{Solutions of the nonlinear system \eqref{eq:nonlin_sys_2} with the exponential cross-immunity kernel. The top (resp. bottom) panels show $I_\asy$ (resp. $\log_{10}\sigma_\asy^2$) versus $m$ and $w$ with $\Ro=2$ (left), $4$ (middle), and $8$ (right). In the top row contours are uniformly distributed in $0.1\leq I_\asy\leq 0.4$ (left), $0.1\leq I_\asy\leq 0.6$ (middle), and $0.1\leq I_\asy\leq 0.7$ (right). The colorbar applies to each row and $\ep=10^{-6}$.}
    \label{fig:full-asy-sweep}
\end{figure}

Fixing $\varepsilon=10^{-6}$ and $\Ro = 2, 4, 8$ we numerically solve \eqref{eq:nonlin_sys_2} for various values of $0<w<1$ and $0<m<1-w$. Specifically, for each value of $w$ we numerically continue the solutions of \eqref{eq:nonlin_sys_2} as $m$ is varied, starting from the large-$m$ asymptotics from \S\ref{subsec:m-regimes}. In Figure \ref{fig:full-asy-sweep} we plot $I_\asy$ and $\sigma_\asy^2$ versus $m$ and $w$ for values of $\Ro=2,4,8$. In each case we observe that $I_\asy$ is monotonically increasing in both $m$ and $w$, so that the endemic burden of infection rises both as infections come to occupy more of the host lifespan and as they become more lethal, with this behavior being more pronounced as $m$ is varied. Unsurprisingly, we also observe that the total proportion of infected individuals increases with the basic reproduction number $\Ro$. This behavior remains qualitatively similar as $\varepsilon$ is varied (not shown).

We interpret the monotonic behavior of $I_\asy$ by recalling that $m =\dimvar{\mu}/(\dimvar{\mu}+\dimvar{\nu}+\dimvar{\omega})$ and $w = \dimvar{\omega}/(\dimvar{\mu}+\dimvar{\nu}+\dimvar{\omega})$. While $m$ captures the ratio of the average infection length to the average lifespan, the parameter $w$ captures the dominance of infection-induced mortality. Larger values of $m$ correspond to a trend towards lifelong infections whereas larger values of $w$ correspond to increasingly fatal infections. The proportion of infections thus increases with both $m$ and $w$, through two distinct effects. Longer lasting infections, i.e.\@ larger values of $m$, increase opportunities for infection and re-infection of the susceptible and recovered classes respectively. On the other hand increasingly fatal infections, i.e.\@ larger values of $w$, reduce the proportion of the recovered class, effectively increasing the \textit{proportion} of infected individuals. While the former mechanism increases the proportion of the infected class through a reduction of the proportion of the susceptible and recovered classes, the latter mechanism targets only a reduction in the size of the recovered class. The results in Figure \ref{fig:full-asy-sweep} indicate that this decrease in the recovered class is sufficient to increase the proportion of the infected class.

The parameter $m$ measures the duration of infection relative to the lifespan of the host, and so distinguishes an acute infection in a long-lived host from one occupying an appreciable fraction of a short life. The parameter $w$ determines how much of the clearance of infection comes from death rather than recovery. Neither parameter depends on the pathogen alone. Both depend on host demography as well, so the same pathogen in two host species can occupy different regions of Figure \ref{fig:full-asy-sweep}. We return to this point in \S\ref{sec:discussion} for influenza A in humans and in wild waterfowl. We also note that because $c_\asy = 2\ep\sqrt{\Ro(1-I_\asy)-1}$ is a decreasing function of $I_\asy$, the contours of $I_\asy$ in the top row of Figure \ref{fig:full-asy-sweep} are also contours of $c_\asy$. Increasing either $m$ or $w$ therefore increases $I_\asy$ and decreases $c_\asy$.

\begin{figure}
    \centering
    \begin{subfigure}{0.5\textwidth}
        \includegraphics[width=1\linewidth]{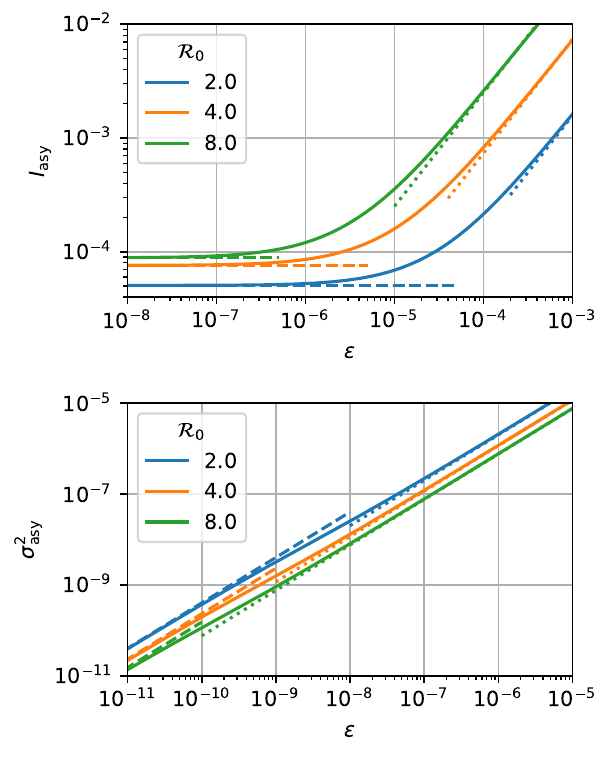}    
    \end{subfigure}%
    \begin{subfigure}{0.5\textwidth}
        \includegraphics[width=1\linewidth]{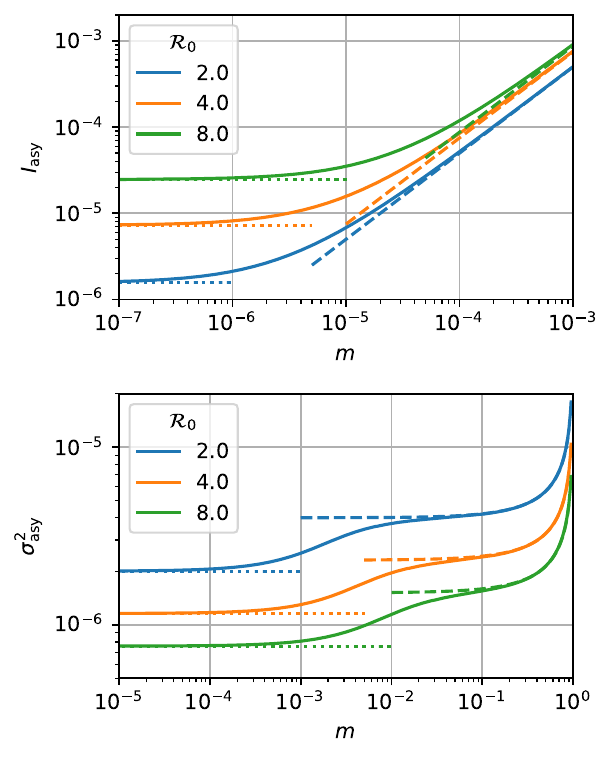}    
    \end{subfigure}
    \caption{Comparison of full numerical simulations of the nonlinear system (solid) with the $m\ll\ep$ (dotted) and $m\gg\sqrt{\ep}$ (dashed) asymptotics. In the left column $m=10^{-4}$ and $w=10^{-2}$ while in the right column $\ep=10^{-6}$ and $w=10^{-8}$.}
    \label{fig:large-small-asy-comparison}
\end{figure}

In Figure \ref{fig:large-small-asy-comparison} we compare the small- and large-$m$ asymptotics of \S\ref{subsec:m-regimes} with the full numerical solutions of the nonlinear system \eqref{eq:nonlin_sys_2}. In each plot the dotted (resp. dashed) curves correspond to the small (resp. large) $m$ asymptotics and are in good quantitative agreement with the solid curves corresponding to the numerical solution of \eqref{eq:nonlin_sys_2}. Note in particular that these figures further support the scalings $I_\asy=O(\ep)$ for $m\ll \ep$, $I_\asy=O(1)$ for $m\gg \sqrt{\ep}$, and $\sigma_\asy^2=O(\ep)$ for all $m$. We remark that parameter values in the original work of Lin et.\@ al.\@ fall squarely in the $m\ll \ep$ regime and are therefore in agreement with the $I_\asy=O(\ep)$ scaling derived here. Human influenza sits well inside this regime, its infections lasting days against a host lifespan of decades. There the prevalence and the antigenic variance depend on the immune evasion accumulated per infection, not on how long infection lasts. Similar quantitative agreement was observed for other choices of $m$ and $\ep$ parameters not included in Figure \ref{fig:large-small-asy-comparison}.

\subsection{Infected and Recovered Distributions}\label{subsec:numerical-comparison}

\begin{figure}[t!]
    \centering
    \includegraphics[width=0.9\linewidth]{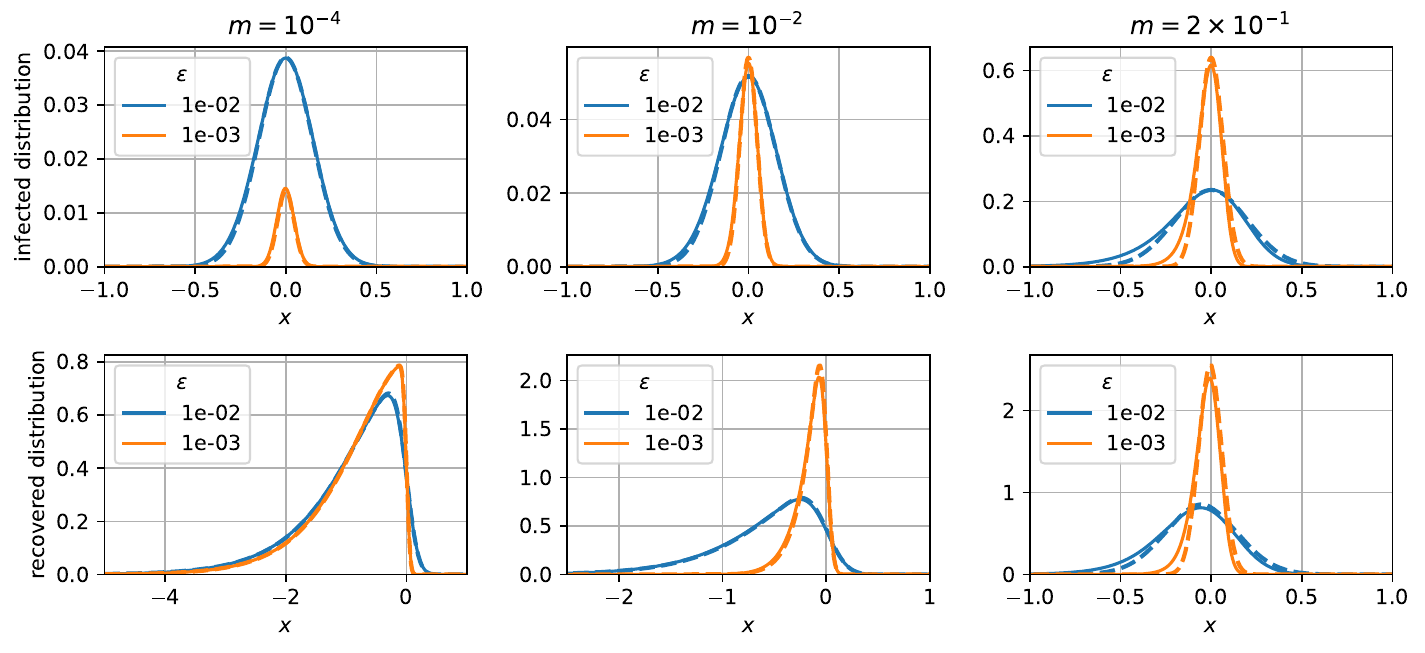}
    \caption{Numerics (solid) and asymptotics (dashed) for $w=10^{-6}$ and $\Ro=2$.}
    \label{fig:num-comp-profile-R0-2}
\end{figure}

\begin{figure}[t!]
    \centering
    \includegraphics[width=0.9\linewidth]{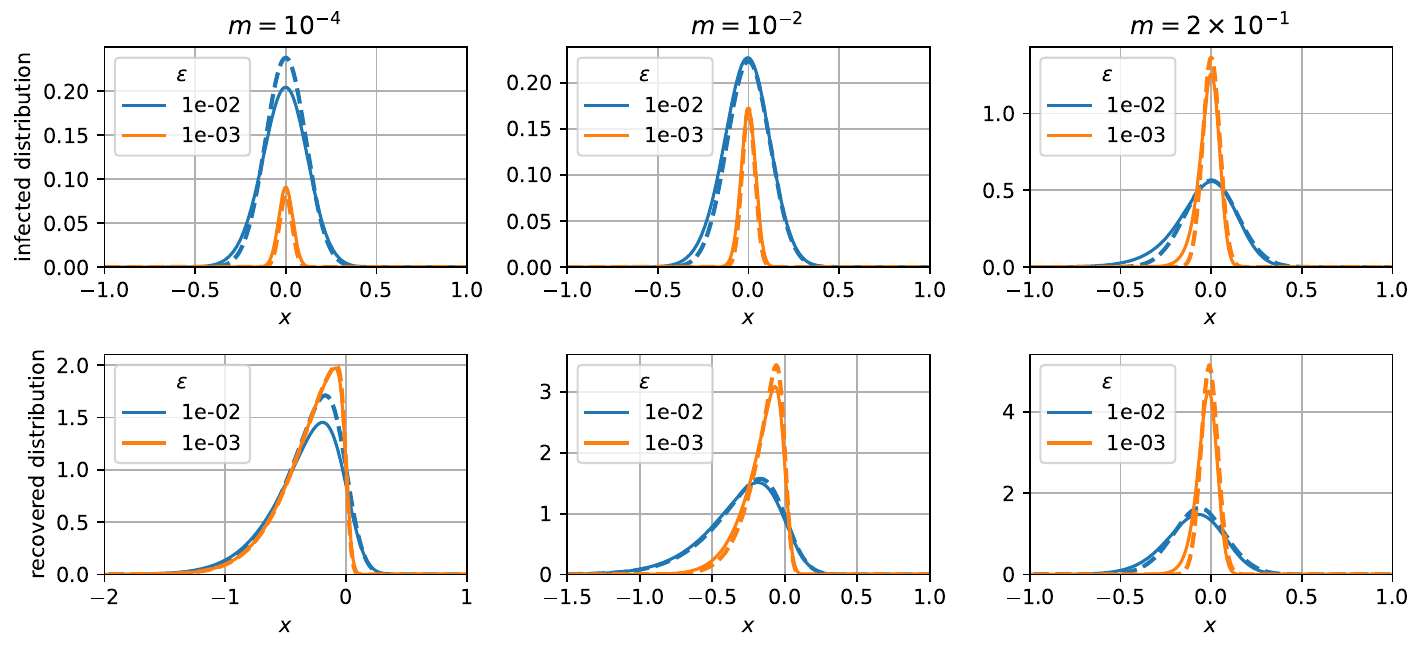}
    \caption{Numerics (solid) and asymptotics (dashed) for $w=10^{-6}$ and $\Ro=4$}
    \label{fig:num-comp-profile-R0-4}
\end{figure}


The asymptotic results in \S\ref{subsec:asymptotics-of-traveling-sol} are validated by comparing with numerical simulations of \eqref{eq:non-dim-sys-normalized}. Restricting to the exponential cross-immunity kernel allows us to localize the nonlocal terms in \eqref{eq:non-dim-sys-normalized} and instead consider the augmented system with \eqref{eq:G1-G2-pdes} as discussed in \S\ref{subsec:exponential-kernel}. We proceed to solve the resulting system of nonlinear differential equations using the SBDF-2 IMEX method \cite{ascher_1995} by considering the truncated domain $0\leq x\leq 70$ with homogeneous Neumann boundary conditions for $i(x,t)$ and Dirichlet boundary conditions for $G_1(x,t)$ and $G_2(x,t)$ informed by \eqref{eq:G1-lim} and \eqref{eq:G2-lim} respectively. We perform numerical simulations for values of $\ep=10^{-2}$ and $\ep=10^{-3}$, for which we use $7000$ and $22000$ uniformly distributed mesh points respectively. The mesh size is informed by the asymptotic theory in order to accurately resolve the profile of the infected distribution which is concentrated on an $O(\sqrt{\ep})$ interval. Simulations were initialized with the asymptotic approximations centered at $x=10$ and evolved over $0\leq t\leq 1500$ with a time-step size of $10^{-3}$. For the remaining parameter values in our numerical simulations we used $m=10^{-4},10^{-2},2\times 10^{-1}$, $w=10^{-6},10^{-3},10^{-1},2\times 10^{-1}$, and $\Ro=2,4,8$.

In all instances numerical solutions settled to traveling pulse solutions after an initial transient. Once the numerical solution settles to a traveling pulse we translate the solution so that the infected distribution attains its maximum at the origin. In Figures \ref{fig:num-comp-profile-R0-2} and \ref{fig:num-comp-profile-R0-4} we compare the resulting numerical solution (solid curves) with the corresponding asymptotic approximations (dashed curves) for select values of $m$, $w$, and $\Ro$. In each instance we see an improvement in the accuracy of the asymptotic solution as $\ep$ is decreased from $10^{-2}$ to $10^{-3}$. However, accuracy of the asymptotic solution deteriorates as either $m$ or $\Ro$ is increased. This behavior is expected since $I$ increases with $m$ and $\Ro$ yet the asymptotic approximation requires that $I$ be sufficiently smaller than $1-1/\Ro$. Estimates of $\Ro$ for influenza A lie in the range $2$ to $5$ \cite{lin_2003}, at the accurate end of the range examined here.

Finally, we comment that our numerical simulations validate the qualitative properties of the infected and recovered distributions deduced from our asymptotic analysis. Specifically, we observe that in all instances the infected distribution is approximately normally distributed. In addition, qualitative properties of the recovered distribution predicted by the large- and small-$m$ asymptotic approximations of \S\ref{subsec:m-regimes} are also borne out in the numerical simulations. Mainly, we observe that $r(x,t)$ is approximately normally distributed for larger values of $m$ but is skewed towards lower values of $x$ for smaller values of $m$.

The infected distribution is unimodal with standard deviation $\sigma_\asy=O(\sqrt{\ep})$, so at any time the infected population occupies an interval of antigenic space of width $O(\sqrt{\ep})$ about $x=ct$. This model behavior reflects the empirical observation of limited standing antigenic diversity of influenza A noted in \S\ref{sec:intro}, and it is a consequence of the analysis rather than an assumption: we imposed only that $i_e>0$ and $i_e\rightarrow0$ as $u\rightarrow\pm\infty$. The recovered distribution attains its maximum at $\overline{u}<0$, so hosts are immune to strains at antigenic positions less than $ct$, and by \eqref{eq:g1'_and_g1_lim} it is the convolution of $r_e$ with $K'$ that makes $g_1$ increasing in $u$. The dependence of the shape of $r_e$ on $m$ follows from \eqref{eq:r_asy_eq_ode_0}: when $m$ is small, $r_e$ decays slowly for $u<0$, since hosts recovered from strains at $u\ll0$ are removed only at rate $m$; when $m$ is large those hosts are removed quickly and $r_e$ is symmetric about a point near $u=0$.


\subsection{Intermediate Virulence Maximizes the Total Number of Infections}\label{subsec:optimal-virulence}

Taking the prevalence of infection as a criterion for viral success, the largest possible infection-induced mortality, or \textit{virulence}, appears optimal (Fig.~\ref{fig:full-asy-sweep}). However, while the total proportion of infections increases with $w$, it does so at the cost of lowering the total population size as seen from \eqref{eq:endemic-N-S}. In particular this trade-off may lead to a decrease in the \textit{total} number of infections. In the remainder of this section we explore this trade-off and show that the total number of infections may instead be maximized for intermediate levels of virulence which depend non-trivially on both demographic and epidemiological parameters.

The trade-off producing this maximum is not the one assumed in the classical theory of virulence evolution described in \S\ref{sec:intro}. There, transmission and virulence are linked by a constraint $\dimvar{\beta}=\dimvar{\beta}(\dimvar{\omega})$, and an interior optimum requires $\dimvar{\beta}''(\dimvar{\omega})<0$ \cite{anderson_1982,alizon_2009}. No such constraint is imposed here, since $\Ro$ and $w$ are independent parameters of \eqref{eq:non-dim-sys-normalized}. The two competing effects are instead that $I_\asy$ increases with $w$, as shown in Figure \ref{fig:full-asy-sweep}, while $N_e$ decreases with $w$ by \eqref{eq:endemic-N-S}. We also recall from \S\ref{sec:intro} that $w$ is a parameter of the model and not an evolving trait, so $w_{\max}$ is the value of $w$ maximizing $I_\tot$ and not the outcome of a selective process.


\begin{figure}[t!]
    \centering
    \includegraphics[width=0.9\linewidth]{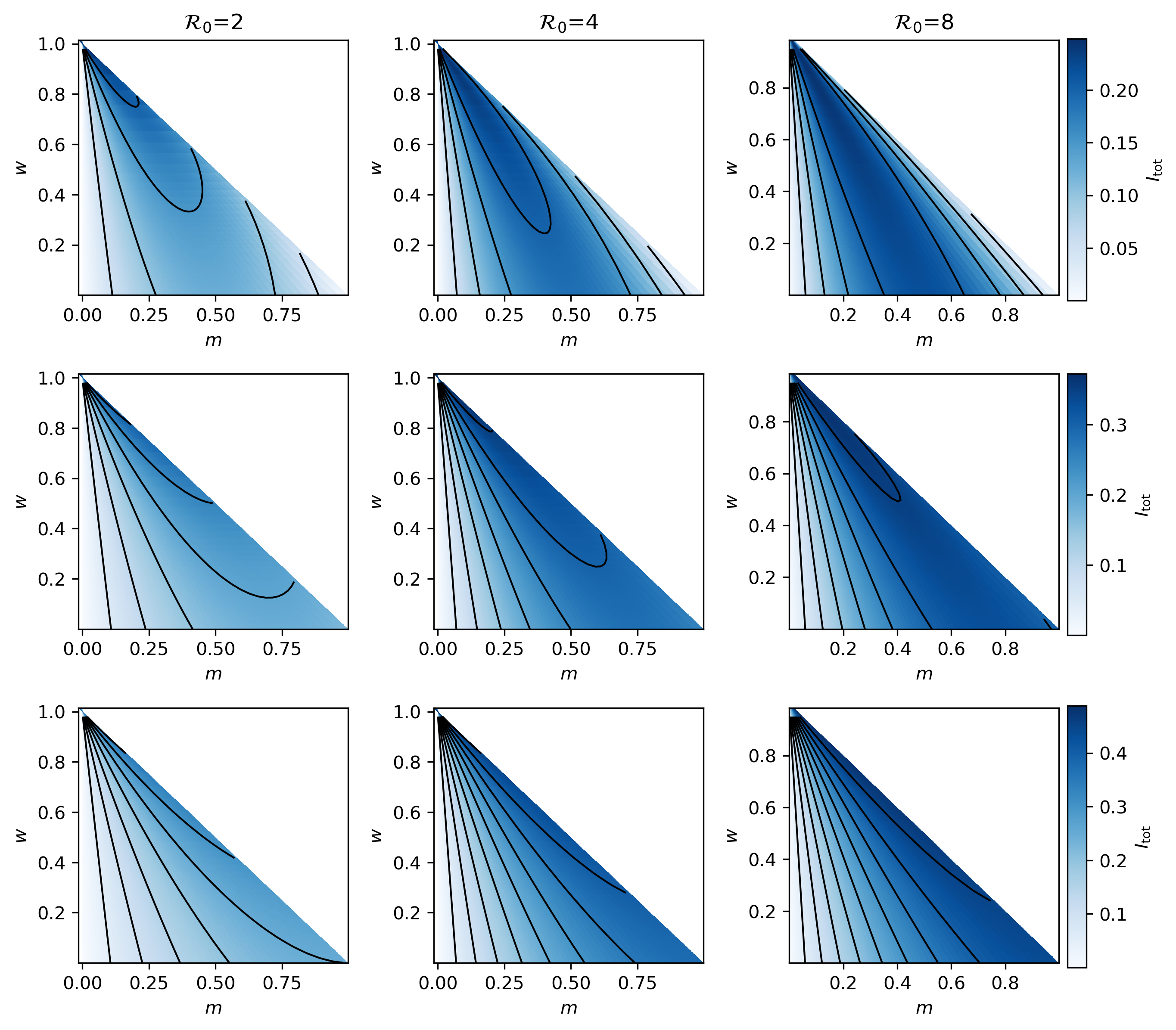}
    \caption{Total infected population $I_\tot=I_\asy N_e$ versus $m$ and $w$ for value of $\Ro=2$ (left), $\Ro=4$ (middle), and $\Ro=8$ (right), and values of $b_0=1.0$ (top), $1.5$ (middle), and $2$ (bottom). Rows share a common colorbar. Contours correspond to evenly distributed values of $0.05\leq I_\tot\leq 0.2$ in the top row, $0.05\leq I_\tot\leq 0.35$ in the middle row, and $0.05\leq I_\tot\leq 0.45$ in the bottom row. The carrying capacity is assumed to be $Q=1$.}
    \label{fig:I_tot-sweep}
\end{figure}

To analyze the consequences of this trade-off, we first note that the total infected population size is given by {\small$I_\tot = I_\asy \left(1 - \frac{m}{b_0} - \frac{w}{b_0}I_\asy\right)Q$.} This expression reveals a non-trivial dependence of the total infected population size on both demographic and epidemiological parameters. In particular, we observe that if $w/b_0$ is sufficiently large then the total infected population may decrease as the proportion increases. Indeed, if the intrinsic birth-rate $b_0$ is sufficiently small then population levels are not replenished fast enough to account for increasing infection-induced fatalities. On the other hand, if $b_0$ is large then the population is replenished quickly rendering the loss from infection-induced fatalities negligible. We therefore anticipate that for small intrinsic birth rates total infections may be maximized at intermediate virulence levels, whereas large values of virulence will maximize total infections when the intrinsic birth rate is sufficiently large. This expectation is borne out in Figure \ref{fig:I_tot-sweep} where we plot $I_\tot$ versus $m$ and $w$ for values of $\Ro=2,4,8$ and $b_0=1,1.5,2$. These plots further illustrate that while the intrinsic birth-rate provides a baseline under which intermediate values of virulence may maximize $I_\tot$, the values of the optimizing virulence depend non-trivially also on the epidemiological parameters $m$ and $\Ro$. We define the \textit{optimal} virulence by
\begin{equation}
    w_{\max} := \mathrm{argmax}_{0\leq w\leq 1-m} I_{\tot},
\end{equation}
and in the remainder of this section we consider first the behavior of $w_{\max}$ in the $m\gg\sqrt{\ep}$ regime and then more generally by numerically solving \eqref{eq:nonlin_sys_2}.

\begin{figure}[t!]
    \centering
    \includegraphics[width=0.5\linewidth]{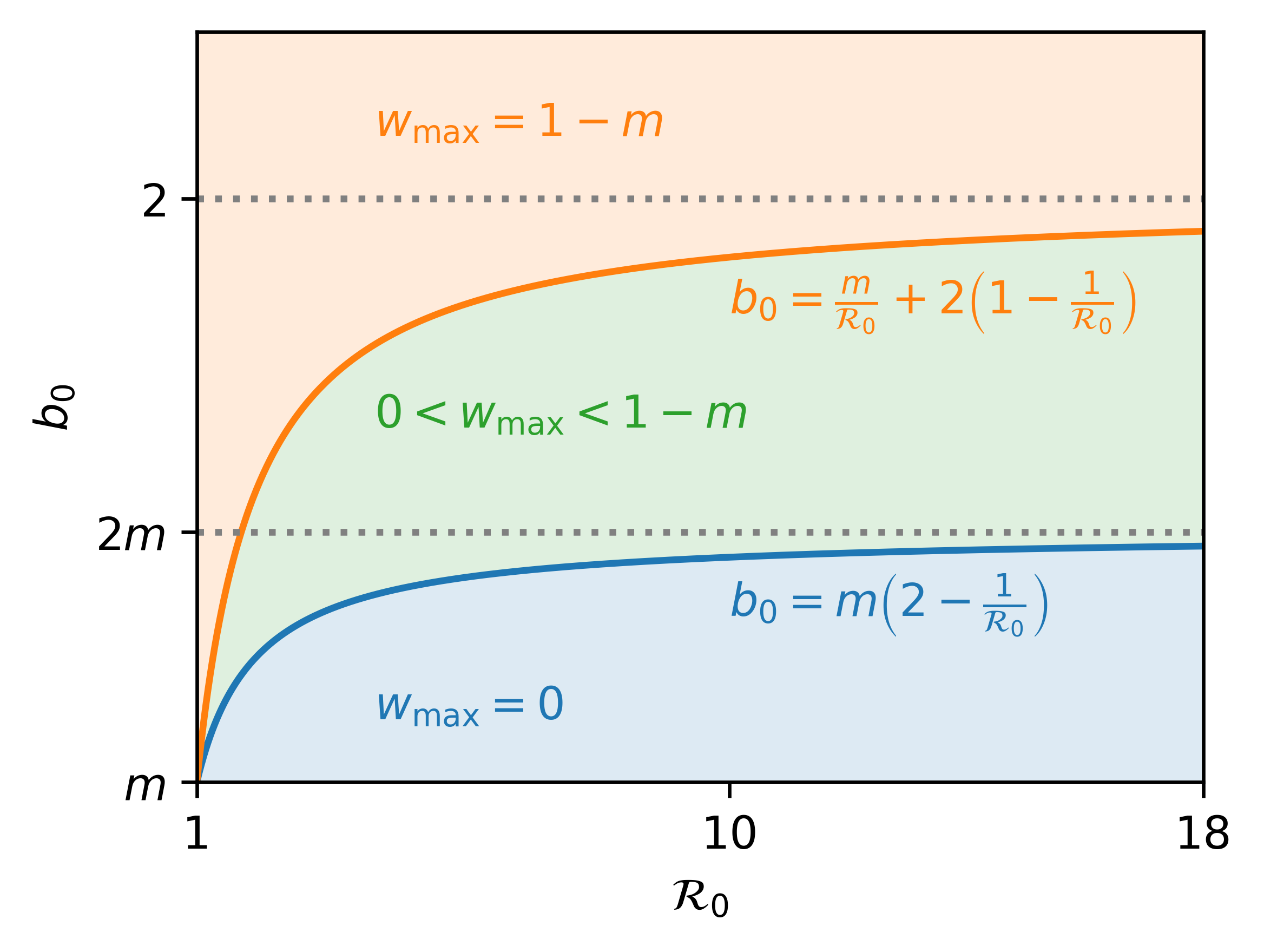} 
    \caption{Optimal virulence $w_\mathrm{max}$ versus $\Ro$ and $b_0$ in the $m\gg\sqrt{\ep}$ regime showcasing three distinct regions where $w_{\max}=1-m$, $w_{\max}$ is intermediate, and $w_{\max}=0$.}
    \label{fig:w_max_large_m_asy}
\end{figure}

In the $m\gg\sqrt{\ep}$ regime considered in \S\ref{subsec:m-regimes} we use \eqref{eq:large-m-results} to obtain the following particularly simple expression for the total infected population
$$
I_\tot \sim \frac{m Q}{(1-w) b_0}\left(1-\frac{1}{\Ro}\right)\left(b_0 - m - \frac{mw}{1-w}\left(1-\frac{1}{\Ro}\right)\right).
$$
Positivity of $I_\tot$ is guaranteed if the intrinsic birth rate is sufficiently high, namely
\begin{equation*}
    b_0 > \frac{m}{1-w}\left(1-\frac{w}{\Ro}\right).
\end{equation*}
It is then straightforward to show that
\begin{equation}\label{eq:large-m-w_max}
    w_{\max} = \begin{cases}
        0, & m<b_0 \leq m\left(2-\frac{1}{\Ro}\right), \\
        \frac{b_0 - m \left(2-\frac{1}{\Ro}\right)}{b_0-\frac{m}{\Ro}}, & m\left(2-\frac{1}{\Ro}\right) < b_0 \leq \frac{m}{\Ro} + 2\left(1-\frac{1}{\Ro}\right), \\
        1-m, & b_0>\frac{m}{\Ro} + 2\left(1-\frac{1}{\Ro}\right)
    \end{cases}
\end{equation}
which we plot in Figure \ref{fig:w_max_large_m_asy} versus $\Ro$ and $b_0$. Intermediate values of virulence are therefore optimal if there is an appropriate  balance between the intrinsic birth rate, the duration of infection, and the basic reproduction number. To explore this balance further we consider each part of \eqref{eq:large-m-w_max} in more detail. Fixing the relative infection duration $m$ and the basic reproduction number $\Ro$, we first see how the intrinsic birth rate measured against the rate at which infections end, $b_0=\dimvar{b}_0/\dimvar{\tau}$, affects the optimal virulence. The three cases in \eqref{eq:large-m-w_max} correspond to $b_0$ small, where $w_{\max}=0$; $b_0$ intermediate, where $0<w_{\max}<1-m$; and $b_0$ large, where $w_{\max}=1-m$ because $N_e$ in \eqref{eq:endemic-N-S} depends only weakly on $w$ once $b_0\gg m+wI_\asy$. If we instead fix the intrinsic birth rate $b_0$ and the relative infection duration $m$, we can see how the basic reproduction number $\Ro$ affects the optimal virulence. Observing that the lower and upper bounds for intermediate virulence in \eqref{eq:large-m-w_max} are monotone increasing in $\Ro$ and respectively asymptote to $2m$ and $2$ as $\Ro\rightarrow\infty$ we deduce the following. First, if $b_0\geq 2$ then $w_{\max}=1-m$ for all $m$ and $\Ro$. If $2m\leq b_0<2$ then the optimal virulence is intermediate for all $\Ro >\frac{2-m}{2-b_0}$, and $w_{\max}=1-m$ otherwise. Finally, if $m<b_0<2m$ then the optimal virulence is intermediate for $(2-m)/(2-b_0)<\Ro<m/(2m-b_0)$, $w_{\max}=1-m$ below this range, and $w_{\max}=0$ above this range. The set of basic reproduction numbers for which virulence is intermediate therefore depends on the intrinsic birth rate in two distinct ways. For $m<b_0<2m$ that set is a bounded interval which widens as $b_0$ increases, its upper endpoint $m/(2m-b_0)$ diverging as $b_0\rightarrow 2m^-$. For $2m\leq b_0<2$ the set is unbounded above and instead contracts as $b_0$ increases, since its lower endpoint $(2-m)/(2-b_0)$ is increasing in $b_0$. Should $\Ro$ fall below this set, then the endemic burden is insufficient to change the overall population size appreciably and $w_{\max}=1-m$. Should $\Ro$ instead rise above it, which is possible only when $m<b_0<2m$, then the effects on the overall population size are sufficiently strong to lower the optimal virulence to $w_{\max}=0$. However, once the intrinsic birth rate grows sufficiently large ($b_0>2$) the population is replenished quickly, so that the optimal virulence is maximal, $w_{\max}=1-m$, for all values of $\Ro$ and $m$.

\begin{figure}
    \centering
    \includegraphics[width=0.9\linewidth]{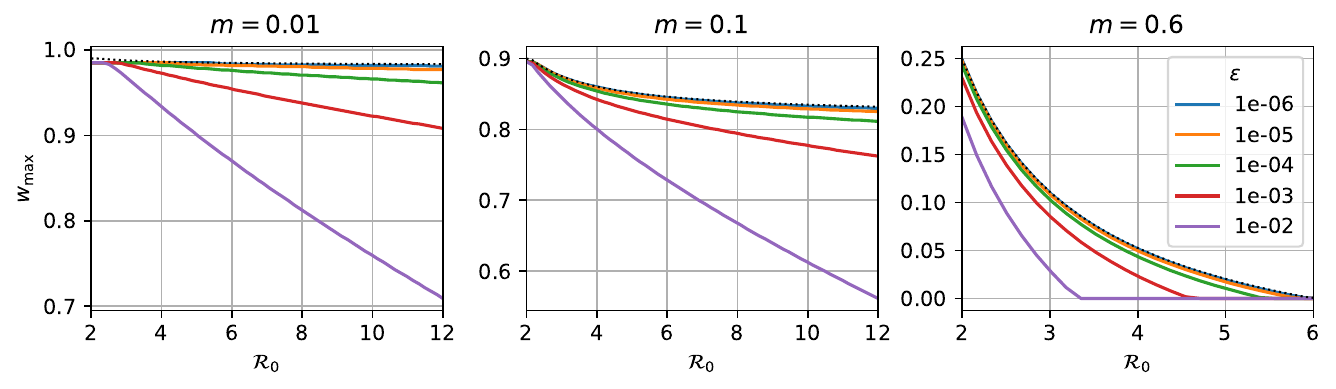}\\ 
    \includegraphics[width=0.9\linewidth]{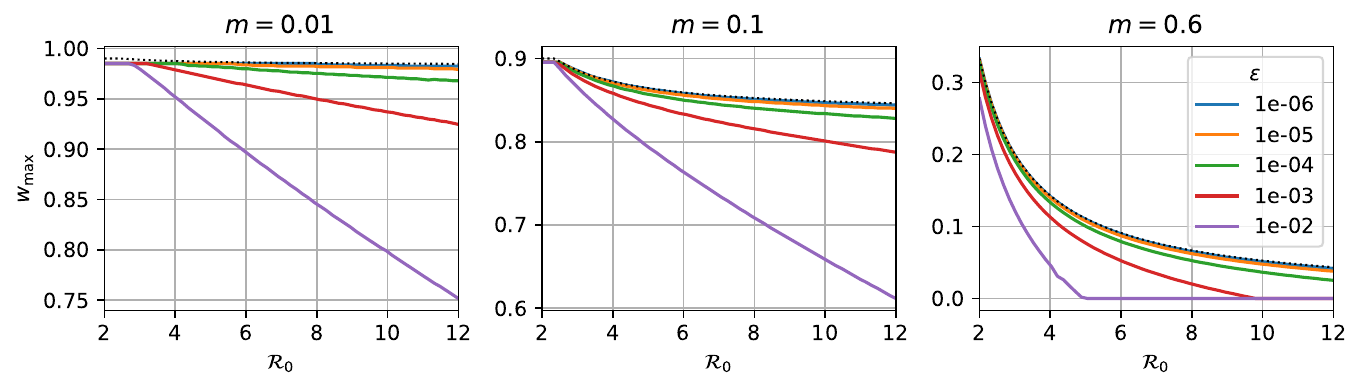}
    \caption{Numerically computed values of $w_\mathrm{max}$ for the exponential cross-immunity kernel for $b_0=1.1$ (top) and $b_0=1.2$ (bottom) and at the indicated values of $m$. Rows share a  common legend shown on the right-most plot.}
    \label{fig:w_max_asy}
\end{figure}

Without the assumption $m\gg \sqrt{\ep}$ we no longer have explicit formulas for $w_{\max}$ though the general heuristic remains true. By first solving the nonlinear system \eqref{eq:nonlin_sys_2} numerically we can then calculate $w_{\max}$ and in Figure \ref{fig:w_max_asy} we plot $w_{\max}$ versus $\Ro$ for values of $b_0=1.1$ (top) and $b_0=1.2$ (bottom) at the indicated values of $m=0.01,0.1,0.6$ and $\ep=10^{-6},10^{-5}, 10^{-4},10^{-3},10^{-2}$. In each case we see that $w_\mathrm{max}$ is a decreasing function of $\ep$ and that it approaches the large $m$ asymptotics as $\ep$ decreases. In particular the $m\gg \sqrt{\ep}$ asymptotics provide an upper bound for $w_{\max}$.

\section{Discussion}\label{sec:discussion}

We have analyzed an eco-evolutionary SIR model in which a pathogen diffuses through a one-dimensional antigenic space and cross-immunity in the recovered class makes the growth rate $g_1(u)$ increasing in $u$. Extending the model of Lin et al.\@ \cite{lin_2003} to include infection-induced mortality, and carrying the asymptotic analysis over a much wider range of parameters, reduces the traveling pulse to two scalars: the prevalence of infection $I_\asy$ and the antigenic variance $\sigma_\asy^2$. Everything else, including the speed of antigenic evolution, follows from these. The reduction makes it possible to ask how epidemiological and demographic parameters shape the rate of antigenic evolution, rather than to compute that rate at one point in parameter space.

We emphasize three results. First, we derive analytically, rather than observe numerically, that the infected distribution is approximately Gaussian, together with the conditions under which the approximation holds; it degrades as $I_\asy$ approaches its upper bound $1-1/\Ro$. Second, $c_\asy\sigma_\asy^2 = 8\ep^2$ in the long-infection regime $m\gg\sqrt{\ep}$, so the speed and the antigenic variance vary inversely at fixed $\ep$ but together when $\ep$ varies. Third, the total number of infected hosts $I_\tot$ can be maximized at an intermediate value of $w$, with no assumed trade-off between transmission and virulence.

\subsection*{Prevalence and the rate of antigenic evolution}

The rate of antigenic evolution depends on the epidemiological state, not on the pathogen alone. The speed {\small$c_\asy = 2\ep\sqrt{\Ro(1-I_\asy)-1}$} decreases with $I_\asy$, where $I_\asy$ solves the system \eqref{eq:nonlin_sys_0}. Any parameter change that raises $I_\asy$ therefore lowers $c_\asy$, so the contours of the top row of Figure \ref{fig:full-asy-sweep} determine both quantities.

That $c_\asy$ decreases with $I_\asy$ runs opposite to a common expectation, namely that more infections supply more replication and hence faster antigenic evolution. In this model $\ep$ fixes the immune evasion accumulated per infection and does not depend on the number of infections. The prevalence enters only through $\Ro(1-I_e)-1 = \Ro(S_e+R_e)-1$, the proportion of hosts that strains at large $u$ can infect. Raising $I_\asy$ lowers that proportion and hence lowers $c_\asy$. Day and Gandon \cite{day_2007} obtain a feedback of the same form in a population-genetic setting, where selection on the pathogen acts with strength proportional to the density of susceptible hosts, so that the epidemiological dynamics enter the evolutionary ones directly.

Within this model, an intervention that lowers $I_\asy$ without altering $\ep$ or $\Ro$ raises the rate of antigenic evolution. We do not press this comparison as a policy claim. The comparison holds between two endemic states rather than describing the transient that follows an intervention, real interventions change several parameters at once, and vaccination in particular alters the cross-immunity kernel $K$ rather than leaving it fixed. But the comparison does show what an eco-evolutionary treatment provides and a purely epidemiological one cannot: $I_\asy$ and $c_\asy$ cannot be adjusted independently.

\subsection*{The rate of antigenic evolution and the standing antigenic variance}

A recurring intuition holds that pathogens with less antigenic variation evolve more slowly. Our results show that this intuition is correct for some comparisons and reversed for others, and that the distinction turns on which parameter separates the two pathogens being compared.

Consider first two host species infected by the same pathogen, influenza A in humans and in wild waterfowl for which $\dimvar{\omega}\approx0$. The parameter that distinguishes them is $m=\dimvar{\mu}/(\dimvar{\mu}+\dimvar{\nu})$, the duration of infection relative to the lifespan of the host. Infection durations are comparable in the two hosts and are short in both: human influenza infections last a few days, and low-pathogenic avian influenza is shed by ducks for roughly ten to twelve days \cite{henaux_2011}. What differs is host lifespan. Taking $\dimvar{\nu}\approx70\,\mathrm{yr}^{-1}$ and $\dimvar{\mu}\approx (70\,\mathrm{yr})^{-1}$ for humans \cite{lin_2003} gives $m\approx 2\times10^{-4}$, whereas a mallard shedding for eleven days with an annual adult survival near one half gives $\dimvar{\nu}\approx34\,\mathrm{yr}^{-1}$, $\dimvar{\mu}\approx0.5\,\mathrm{yr}^{-1}$, and $m\approx1.5\times10^{-2}$. The parameter is thus roughly two orders of magnitude larger in waterfowl, and the difference comes from host demography rather than from any property of the infection. We stress that avian influenza infections are not long-lasting in absolute terms. Ducks clear infection and acquire immunity. The ratio, not the duration of infection, is what differs.

Human influenza therefore sits in the short-lived infection regime of \S\ref{subsec:m-regimes}, in which the prevalence and the antigenic variance are independent of $m$ to leading order, while influenza in waterfowl sits at or above the crossover into the long-infection regime of \S\ref{subsec:m-regimes}. Which side of the crossover $m\sim\sqrt{\ep}$ the avian case falls on depends on $\ep$ as well as on demography. The relevant $\ep$ is plausibly smaller in birds than in humans, since immune selection appears weaker and cross-immunity broader in the avian reservoir \cite{webster_1992}. A smaller $\ep$ lowers the crossover and places the avian estimate above it. Our numerical solution of the full nonlinear system, plotted in Figure \ref{fig:large-small-asy-comparison}, shows that $I_\asy$ and $\sigma_\asy^2$ leave their small-$m$ plateaux and begin to increase once $m$ exceeds $\ep$, so the qualitative distinction does not require the closed-form expressions \eqref{eq:large-m-results} to hold exactly.

Our results then account for two features of influenza A in its wild bird reservoir. The first is evolutionary stasis: amino acid changing substitutions accumulate slowly and haemagglutinin antigenicity is nearly constant over decades \cite{webster_1992}, even though nucleotide substitution itself is not slow \cite{chen_2006}. The second is the greater antigenic diversity maintained at any one time. Both follow from the long-infection asymptotics, in which a larger $m$ raises $\sigma_\asy^2$ and lowers $c_\asy$. We add one caveat on magnitudes and one on the nature of the diversity. On magnitudes, the $m$-driven reduction in speed is proportional to $\sqrt{1-m-w}$ by \eqref{eq:large-m-results}, so it becomes appreciable only when $m$ is an appreciable fraction of unity. A tenfold contrast in the rate of antigenic evolution is more naturally attributed to a tenfold difference in $\ep$, on which $c_\asy$ depends linearly. We would therefore expect the difference in $m$ to account for the difference in standing antigenic variance more convincingly than for the difference in rate. On diversity, much of the diversity of influenza A in waterfowl is diversity of haemagglutinin and neuraminidase subtype, with sixteen HA and nine NA subtypes co-circulating and reassorting \cite{diskin_2020}. Subtype diversity is a form of diversity the present model does not represent, since it is not variance along a single antigenic axis, and our prediction concerns the latter.

The second comparison is between two pathogens in the same host, influenza A and measles in humans. Here the separating parameter is $\ep$, which measures the immune evasion a lineage accumulates over one average infection. Measles is antigenically stable, and in our terms $\ep$ is very small. Because $c_\asy = O(\ep)$ and $\sigma_\asy^2 = O(\ep)$, the model predicts that measles should show both slower antigenic evolution and lower standing antigenic variance than influenza, which is what is observed. The contrast with the previous comparison is the point: antigenic variance and rate of evolution are positively related across pathogens that differ in mutational input, and inversely related across host species that differ in the duration of infection relative to lifespan.

Our finding that longer infections permit greater antigenic variance complements an earlier result of Gog and Grenfell \cite{gog_2002}. They studied a discrete many-strain model with polarized immunity, and found that short infectious periods relative to host lifetime produce a single cluster of strains moving through antigenic space, whereas prolonged infections allow several clusters to coexist. We reach the same conclusion by a different route, through an asymptotic reduction of a nonlocal partial differential equation rather than a status-based discrete-strain formulation. The agreement suggests that the effect follows from cross-immunity itself and not from a particular formalism.

\subsection*{Virulence}

The classical theory of virulence evolution predicts an interior optimum only under an assumed constraint. If transmission and virulence are linked by $\dimvar{\beta}=\dimvar{\beta}(\dimvar{\omega})$, then selection maximizes $\Ro=\dimvar{\beta}(\dimvar{\omega})/(\dimvar{\mu}+\dimvar{\nu}+\dimvar{\omega})$ and the optimum is interior when $\dimvar{\beta}''(\dimvar{\omega})<0$ \cite{anderson_1982,alizon_2009}. The intermediate optimum we report in \S\ref{subsec:optimal-virulence} arises without such a constraint. Transmission is independent of virulence in our model, and $\Ro$ and $w$ are independent parameters. What produces the optimum instead is a demographic feedback. Raising $w$ increases the \textit{proportion} of hosts infected, because it removes recovered individuals and so shrinks the immune pool, but it simultaneously lowers the endemic population size $N_e$ through \eqref{eq:endemic-N-S}. The total number of infections is the product of these two quantities, and the product is maximized at an interior value of $w$ whenever the intrinsic birth rate $b_0$ is too small to replenish the hosts that virulence removes. The optimum is thus set by a balance between epidemiology and host demography, not by a physiological trade-off within the pathogen.

Virulence is a fixed parameter in our model, not an evolving trait. The quantity $w_{\max}$ maximizes $I_\tot$, but it is not the outcome of a selective process, and we do not claim that selection drives $w$ toward $w_{\max}$. Establishing a claim of that kind requires a model in which pathogen lineages differ in virulence and compete, which ours does not. Our result identifies which pathogens, among a set differing in virulence, would sustain the largest number of infections in an endemic state.

\subsection*{Fitness and the rate of adaptation}


In \S\ref{subsec:identities} we derive three identities holding exactly for the traveling pulse: the average growth rate across circulating variants vanishes, the speed of antigenic advance equals the covariance between a variant's growth rate and its antigenic position, and the variance in growth rate balances the rate at which the growth rate declines at each fixed antigenic position. The condition on which our asymptotic construction closes, $g_1'(0)\sigma_\asy^2=c_\asy$, is the leading-order form of the covariance identity. Written as \eqref{eq:fisher-balance}, the variance identity is a statement of Fisher's fundamental theorem, with one qualification. Fisher's theorem concerns the change in average fitness at a fixed fitness function, whereas here the growth rate at each antigenic position changes in time (as the recovered population changes its composition), and the term $-\langle\partial_tG_1\rangle$ is what Fisher called the deterioration of the environment. That term is not exogenous in our model. The recovered distribution determining $G_1$ is produced by the pathogen's own past infections, so the infections already produced lower the growth rate at exactly the rate at which selection raises it. Mustonen and L\"assig \cite{mustonen_2010} treat adaptation under a fitness function that varies in time, and the notion of a fitness seascape, meaning a fitness function that changes as the population evolves, has been applied to influenza in predictive models \cite{luksza_2014}. What distinguishes the present model is that the seascape is generated by the solution itself. We note also that $G_1$ is frequency dependent, so while the identity holds, the classical conclusion that selection increases mean fitness does not: the average growth rate here is constant at zero.

\subsection*{Limitations and extensions}

We assume the same one-dimensional antigenic space of Lin et al.\@ \cite{lin_2003}, and it remains the strongest assumption in our model. The trunk-dominated genealogy of human influenza A/H3N2 motivates this choice \cite{smith_2004,bedford_2014}, but haemagglutinin carries several antigenic sites that can change independently, so escape is not confined to one direction, a priori. Reassortment between co-circulating lineages moreover produces discontinuous antigenic change, which no classical diffusion process describes \cite{nelson_2007}. Both features require a higher-dimensional antigenic space, in which the traveling pulse would give way to a moving distribution with its own shape dynamics.

We stress that the trunk-dominated genealogy is a property of the human reservoir, not of influenza A generally, and this distinction bears on the comparison with waterfowl made above. The avian reservoir shows the opposite pattern: many subtypes co-circulate and reassort \cite{webster_1992,diskin_2020}, with no single lineage from which successive variants descend. The subtype diversity we invoked in that comparison is therefore not the antigenic variance $\sigma_\asy^2$ of our model, and the two should not be conflated. The one-dimensional idealization consequently applies less comfortably to waterfowl than to humans. We intend the comparison between the two hosts as a comparison of $m$ within the model, and not as a claim that a single antigenic axis describes the avian reservoir. The ordering embedded in the cross-immunity kernel, under which recovery from a strain confers complete immunity to all earlier strains, likewise has no counterpart in two dimensions.

A different treatment of antigenic variation retains discrete strains and tracks immune status rather than immune history, which keeps the state space linear in the number of strains \cite{gog_2002}. That approach and ours have complementary strengths. The discrete formulation accommodates arbitrary antigenic geometry, while the continuum formulation admits the asymptotic reduction that makes the dependence on parameters explicit.

We treat the dynamics deterministically throughout, which matters most for the speed. As noted in \S\ref{sec:intro}, the leading edge of the pulse selects $c_\asy$, and there the infected density is small. Pulled fronts are sensitive to the far tail of the distribution, precisely where a deterministic description of a finite population is least accurate. The discreteness of hosts imposes an effective cutoff that lowers the speed by a correction decaying only as the inverse square of the logarithm of population size \cite{brunet_1997}. Lin et al.\@ \cite{lin_2003} estimated this correction at a few percent for a population of $10^6$, small relative to the effects we report but decaying too slowly to vanish for realistic populations. Because the correction always lowers the speed, our expressions give upper bounds on the rate of antigenic evolution. A stochastic treatment would also admit questions our model cannot pose, including extinction of the pathogen and the fate of the transient side branches that the deterministic pulse suppresses.

Our results also connect to the literature on the rate of adaptation in asexual populations, where a localized distribution in fitness advances at a speed determined by the mutational input and the strength of selection \cite{tsimring_1996,desai_2007,rouzine_2008}. Our $\ep$ corresponds to the mutational input in those models, and our $\sigma_\asy^2$ to the variance of the advancing distribution. The relation $c_\asy\sigma_\asy^2 = 8\ep^2$ is of the same type as the speed-variance relations obtained there. The models differ in one respect, identified in \S\ref{sec:intro}. Those models specify the fitness gradient externally, so the speed depends on parameters alone. Here the gradient $g_1'(0)$ is a functional of the recovered distribution, which the pathogen itself produces, and which we must solve for simultaneously. This difference is the content of the eco-evolutionary feedback described above, and it is why $I_\asy$ appears in the expression for $c_\asy$.

A natural extension would let virulence evolve alongside antigenic type. Our model treats virulence as fixed. A pathogen carrying both an antigenic position and a virulence, with lineages differing in both, would occupy a two-dimensional trait space. Cross-immunity would generate frequency-dependent selection on the first coordinate, and classical virulence theory would describe selection on the second. The population-genetic formulation of Day and Gandon \cite{day_2007,day_2012} suits this problem, since it tracks the joint distribution of traits and the covariances between them rather than an evolutionarily stable endpoint. Their formulation also does not assume that epidemiological dynamics are fast relative to evolutionary ones, an assumption untenable here because the two proceed on a common time scale by construction. Such a model would connect to the literature on multiple infection and within-host competition \cite{van_1995,alizon_2013,alizon_2008decreased}, where the classical prediction of an interior virulence optimum is most often overturned. Such a model would also connect to the finding of Griette et al.\@ \cite{griette_2015} that virulence at a spreading front differs from virulence behind it, though our traveling pulse is a spreading front in antigenic rather than geographic space.

Most of evolutionary theory takes the environment as given and asks how a population adapts to it. Most of epidemiology takes the pathogen as given and asks how a population responds to it. Neither holds for an antigenically evolving pathogen. The pathogen's own earlier infections produce the recovered distribution that selects it, so the selective environment is an integral of the evolutionary history. The rate at which the pathogen escapes that immunity in turn determines how many hosts it infects, so the epidemiological state is an outcome of the evolutionary rate. Evolution and ecology are two descriptions of one process here, not two processes on separate time scales. The traveling pulse is the object in which they coincide, and our reduction shows that all of it, the profiles, the prevalence, the antigenic variance, and the speed, is fixed by two numbers. That tractability, more than any single result it yields, is the case for treating antigenic drift with nonlocal partial differential equations.

\appendix 

\section{Asymptotics in the $m\gg\sqrt{\ep}$ Regime}\label{app:asymptotics-large-m}

The explicit expression \eqref{eq:r_asy_pos} for $r_\asy(u)$ when $u\geq 0$ hints at the distinguished regime $m\gg \sqrt{\ep}$. Indeed, using the asymptotics $\erfc(z)\sim \pi^{-1/2}z^{-1}e^{-z^2}$ as $z\rightarrow+\infty$ we deduce that
\begin{equation*}
    r_\asy(u) \sim \frac{1-m-w}{m(1 + \frac{c_\asy}{\sigma_\asy^2 m}u)}\frac{I_\asy}{\sigma_\asy\sqrt{2\pi}}\exp\left(-\tfrac{u^2}{2\sigma_\asy^2} \right) \sim \frac{1-m-w}{m}\frac{I_\asy}{\sigma_\asy\sqrt{2\pi}}\exp\left(-\tfrac{u^2}{2\sigma_\asy^2} \right),
\end{equation*}
which suggests that $r_\asy(u)$ may be normally distributed. To investigate this for $u<0$ we seek a solution of the form \eqref{eq:r_asy_rescale_0}. Since $m\gg \sqrt{\ep}$ the asymptotics $\erfc(z)\sim \pi^{-1/2}z^{-1}e^{-z^2}$ as $z\rightarrow+\infty$  lead us to deduce that \eqref{eq:r_asy_rescale_1a} becomes
\begin{equation*}
    \rho'(u) - \left(\tfrac{\Ro I_\asy}{c_\asy}K(-u) + \tfrac{m}{c_\asy}\right)\rho(u) = -\tfrac{m}{c_\asy} \exp\left(-\tfrac{u^2}{2\sigma_\asy^2}\right),\qquad u<0.
\end{equation*}
We claim, and for the moment assume, that in this parameter regime $I_\asy=O(1)$, and $\sigma_\asy^2=O(\ep)$. Seeking first an inner solution we let $u = \sigma_\asy U$ and write $\rho(\sigma_\asy U) = P(U)$ to get the inner problem
\begin{equation*}
    \tfrac{1}{\sigma_\asy}P'(U) - \left(\tfrac{\sigma_\asy \Ro I_\asy}{c_\asy } U + \tfrac{m}{c_\asy }\right)P(U) = -\tfrac{m}{c_\asy} e^{-U^2/2},\quad U<0;\qquad P(0) = 1.
\end{equation*}
Since $m \gg \sqrt{\ep}$ we deduce $P(U)\sim e^{-U^2/2}$. On the other hand, we can write the outer solution as
\begin{equation*}
    \rho(u) \sim \frac{m}{m + \Ro I_\asy K(-u)}\exp\left(-\tfrac{u^2}{2\sigma_\asy^2}\right),
\end{equation*}
which also captures the inner solution.

Using $K(-u)\equiv 0$ for $u\geq 0$ we combine the $u<0$ and $u\geq 0$ asymptotics to get
\begin{equation}\label{eq:r_asy_pos_asy}
    r_\asy(u) \sim \frac{1-m-w}{m + \Ro I_\asy K(-u)}\frac{I_\asy}{\sigma_\asy\sqrt{2\pi}}\exp\left(-\frac{u^2}{2\sigma_\asy^2}\right).
\end{equation}
The recovered population is therefore approximately normally distributed and localized so
\begin{equation*}
        \int_0^\infty K(u)r_\asy(-u)du \sim \frac{\sigma_\asy}{\sqrt{2\pi}}\frac{1-m-w}{m}I_\asy,\qquad \int_0^{\infty} K'(u)r_\asy(-u)du \sim \frac{1-m-w}{2 m}I_\asy,
\end{equation*}
with which \eqref{eq:nonlin_sys_0} simplifies to
\begin{align*}
    & \frac{\sigma_\asy}{\sqrt{2\pi}}\frac{1-m-w}{m}I_\asy + \frac{m + w I_\asy}{m + \Ro I_\asy} - \frac{1}{\Ro} = 0, \\
    & \sigma_\asy^2\Ro \frac{1-m-w}{2m}I_\asy - c_\asy  = 0. 
\end{align*}
Solving yields the leading order asymptotics in \eqref{eq:large-m-results}. Note that the asymptotic approximation is consistent with the assumptions $I_\asy=O(1)$ and $\sigma_\asy^2=O(\ep)$ provided that
\begin{equation}
    (\Ro - 1)\frac{1-m-w}{1-w} = O(1).
\end{equation}
In particular the accuracy of the asymptotic approximation deteriorates when this condition is not satisfied, e.g. if $0\leq 1-m-w\ll 1$, $0\leq 1-w\ll 1$, or $0\leq \Ro - 1 \ll 1$.


\section{Asymptotics in the $\ep^2\ll m\ll \ep$ Regime}\label{app:asymptotics-small-m}

in this regime we claim, and for the moment assume, that $I_\asy=O(\ep)$ and $\sigma_\asy^2 =O(\ep)$. Numerical calculation of solutions to the nonlinear system \eqref{eq:nonlin_sys_0} suggest that in fact the results from this section are valid for all $m\ll\ep$, though the constraint $m\gg \ep^2$ is here needed for a technical reason.  We seek a solution of the form \eqref{eq:r_asy_rescale_0} and since $m\ll \ep$ we find that \eqref{eq:r_asy_rescale_1a} becomes
\begin{equation}\label{eq:m-small-asy-r-ode}
    \rho'(u) - \left(\tfrac{\Ro I_\asy}{c_\asy}K(-u) + \tfrac{m}{c_\asy}\right)\rho(u) = -\tfrac{1}{\sigma_\asy}\sqrt{\tfrac{2}{\pi}}\exp\left(-\tfrac{u^2}{2\sigma_\asy^2}\right),\quad u<0.
\end{equation}
Let $u=\sqrt{2}\sigma_\asy U$ and seek an inner solution of the form $\rho(\sqrt{2}\sigma_\asy U) \sim P(U)$ satisfying
\begin{equation*}
    P'(U) - \sqrt{2}\sigma_\asy\left(-\tfrac{\sqrt{2}\Ro I_\asy \sigma_\asy}{c_\asy}U + \tfrac{m}{c_\asy} \right)P(U) = -\tfrac{2}{\sqrt{\pi}}e^{-U^2},\quad U<0.
\end{equation*}
The dominant balance is between the first and last terms and therefore the leading order inner solution for $u=O(\sigma_\asy)$ is
\begin{equation*}
    \rho(u) \sim \erfc\left(\tfrac{u}{\sigma_\asy\sqrt{2}}\right).
\end{equation*}
On the other hand, if $|u|\gg \sigma_\asy$ the right-hand-side of \eqref{eq:m-small-asy-r-ode} is negligible and therefore
\begin{equation*}
    \rho(u) \sim 2\exp\left(\tfrac{m}{c_\asy}u - \tfrac{\Ro I_\asy}{c_\asy}\int_0^{-u}K(z)dz\right),
\end{equation*}
where the multiplicative factor of $2$ comes from matching with the far-field behavior of the inner solution. We thus have the following composite asymptotic solution for all $u<0$
\begin{equation}\label{eq:m-small-asy-r-comp}
    r_\asy(u) \sim \frac{1-m-w}{c_\asy}I_\asy\left( e^{\frac{m}{c_\asy}u - \frac{\Ro I_\asy}{c_\asy}\int_0^{-u}K(z)dz} + \frac{1}{2}\erfc\left(\frac{u}{\sigma_\asy\sqrt{2}}\right) - 1 \right).
\end{equation}

Using the composite asymptotics we can approximate the integrals appearing in \eqref{eq:nonlin_sys_0}. First, integrating by parts we obtain
\begin{align*}
    \int_0^\infty K(u)\exp\left(-\tfrac{m}{c_\asy}u -  \tfrac{\Ro I_\asy}{c_\asy}\int_0^uK(z)dz\right)du \sim \tfrac{c_\asy}{\Ro I_\asy} - \tfrac{m}{\Ro I_\asy}\mathcal{K}\left(\tfrac{\Ro I_\asy}{c_\asy}\right),
\end{align*}
where $\mathcal{K}(\cdot)$ is defined in \eqref{eq:script-K-H-def}. On the other hand
\begin{equation*}
    \int_0^\infty K(u)\left(\tfrac{1}{2}\erfc\left(-\tfrac{u}{\sigma_\asy\sqrt{2}}\right) - 1\right)du \sim \sigma_\asy^2\int_0^\infty z\left(\erfc(-z)-2\right)dz = -\tfrac{1}{4}\sigma_\asy^2.
\end{equation*}
Expanding in powers of $m$ the nonlinear equation \eqref{eq:nonlin_sys_0_1} becomes
\begin{equation*}
    -\tfrac{m}{\Ro c_\asy}\left(\tfrac{c_\asy}{\Ro I_\asy}(\Ro -w) - (1-w)\mathcal{K}\left(\tfrac{\Ro I_\asy}{c_\asy}\right)  - c_\asy + \tfrac{I_\asy \sigma_\asy^2}{4\Ro}  \right) - (1-w)\tfrac{I_\asy\sigma_\asy^2}{4c_\asy} + \mathrm{h.o.t} = 0,
\end{equation*}
where $\mathrm{h.o.t.}$ indicates \textit{higher order terms}. Since we are assuming that $I_\asy,\sigma_\asy^2=O(\ep)$ and $\ep^2\ll m\ll \ep$, the leading order problem becomes
\begin{equation*}
    \zeta\mathcal{K}\left(\zeta\right) - \frac{\Ro - w}{1-w} =0,
\end{equation*}
where $\zeta = \Ro I_\asy / c_\asy$. This expression can be solved for a unique positive value $\zeta=\zeta_\star$ depending only on  the cross-immunity kernel and the ratio $(\Ro -w)/(1-w)$. This can be seen by writing
$$
\zeta\mathcal{K}(\zeta) = \int_0^\infty\exp\left(-\int_0^uK(\zeta^{-1}z)dz \right)du,
$$
which we see is monotone increasing and tends to $1$ as $\zeta\rightarrow 0^+$. Therefore $I_\asy = \tfrac{c_\asy \zeta_\star}{\Ro}$ and since
\begin{equation*}
    c_\asy = 2\ep\sqrt{\Ro(1-I_\asy)-1} \sim 2\ep\sqrt{\Ro\left(1-\tfrac{c_\asy \zeta_\star}{\Ro}\right)-1}\sim 2\ep\sqrt{\Ro-1},
\end{equation*}
we deduce that $I_\asy=O(\ep)$, which is consistent with our ongoing assumption.

To determine $\sigma_\asy$ we only need the leading order asymptotics of the integral appearing in \eqref{eq:nonlin_sys_0_2}, namely
\begin{equation*}
    \int_0^\infty K'(u) r_\asy(-u)du\sim \tfrac{1-w}{c_\asy}\mathcal{H}\left(\tfrac{\Ro I_\asy}{c_\asy}  \right) + O(\sigma_\asy,m),
\end{equation*}
where $\mathcal{H}(\cdot)$ is defined in \eqref{eq:script-K-H-def}. With this \eqref{eq:nonlin_sys_0_2} becomes
\begin{equation*}
    (1-w)\sigma_\asy^2 \tfrac{\Ro I_\asy}{c_\asy} \mathcal{H}\left(\tfrac{\Ro I_\asy}{c_\asy}\right)  - c_\asy = 0,
\end{equation*}
which we can use to solve for $\sigma_\asy^2$, finding in particular that $\sigma_\asy^2=O(\ep)$. We have thus established \eqref{eq:small-m-asymptotics}.

We conclude by using the composite asymptotic approximation \eqref{eq:m-small-asy-r-comp} to determine an approximation for the value of $u$ at which $r_\asy(u)$ attains its maximum. Setting the derivative of \eqref{eq:m-small-asy-r-comp} to zero gives
\begin{equation*}
    \left(\tfrac{m}{c_\asy} + \tfrac{\Ro I_\asy}{c_\asy}K(-u)\right)\exp\left(\tfrac{u^2}{2\sigma_\asy^2} + \tfrac{m}{c_\asy}u - \tfrac{\Ro I_\asy}{c_\asy}\int_0^{-u}K(z)dz\right) - \tfrac{1}{\sigma_\asy\sqrt{2\pi}} = 0
\end{equation*}
We anticipate $r_\asy(u)$ to peak for a small value of $u<0$ so that only the first term in the exponential dominates. Now let $u = -\sigma_\asy \sqrt{2} z$ with $z\ll \sigma_\asy^{-1}$ to get the approximation
\begin{equation}\label{eq:m-small-peak-eq}
    \left(\tfrac{m}{c_\asy} + \sqrt{2}\tfrac{\Ro I_\asy \sigma_\asy}{c_\asy}z   \right) e^{z^2} = \tfrac{1}{\sigma_\asy\sqrt{2\pi}}.
\end{equation}
There are here two possible scalings depending on whether the first or second term in the brackets dominates. If $m/c_\asy \gg \sigma_\asy z$ then to leading order
\begin{equation*}
    z\sim \sqrt{\log\left(\tfrac{c_\asy}{m\sigma_\asy\sqrt{2\pi}}\right)},
\end{equation*}
which requires that $\ep^{3/2}\sqrt{|\log\ep|}\ll m \ll \ep $ for the scaling to be consistent. If instead $m/c_\asy \ll \sigma_\asy z$ we square both sides of \eqref{eq:m-small-peak-eq} to get
\begin{equation*}
    2z^2 e^{2z^2} \sim \left(\tfrac{c_\asy}{\Ro I_\asy \sigma_\asy^2 \sqrt{2 \pi}}\right)^2,
\end{equation*}
which we can solve in terms of the Lambert-W function $\mathcal{W}(\cdot)$. Noting the large argument asymptotics of the Lambert-W function \cite{NIST:DLMF} we find that 
\begin{equation*}
    z \sim \sqrt{\tfrac{1}{2}\mathcal{W}\left(\tfrac{1}{2\pi\sigma_\asy^4}\left(\tfrac{c_\asy}{\Ro I_\asy}\right)^2  \right)} \sim \sqrt{2\log\left(\tfrac{1}{\sigma_\asy}\right)},
\end{equation*}
which  requires $\ep^2 \ll m \ll \ep^2\sqrt{|\log\ep|}$ for the scaling to be consistent. In summary we find that $r_\asy(u)$ attains its maximum at 
\begin{equation}\label{eq:m-small-peak-sep}
    \overline{u} \sim \begin{cases}
        -\sigma_\asy\sqrt{\mathcal{W}\left(\tfrac{c_\asy^2}{2\pi \Ro^2 I_\asy^2 \sigma_\asy^4} \right)}, & \ep^2\ll m \ll \ep^2\sqrt{|\log\ep|}, \\
        -\sigma_\asy\sqrt{2\log\left(\tfrac{c_\asy}{m\sigma_\asy\sqrt{2\pi}}  \right)}, & \ep^{3/2}\sqrt{|\log\ep|}\ll m \ll \ep.
    \end{cases}
\end{equation}
In particular, we note that in both cases $\overline{u} = O (\sqrt{\ep|\log\ep|})$ and is negative.

\section{The Nonlinear System for the Exponential Cross-Immunity Kernel}\label{app:exponential-kernel}

In this appendix we use the exponential cross-immunity kernel \eqref{eq:exponential-kernel} to develop an iterative method for evaluating the integrals appearing in \eqref{eq:nonlin_sys_0}. In addition we explicitly calculate the functions $\mathcal{K}(\cdot)$ and $\mathcal{H}(\cdot)$ defined in \eqref{eq:script-K-H-def} appearing in the $\ep^2\ll m\ll\ep$ regime.

For integers $n\geq 0$ let
\begin{equation*}
    \mathcal{Q}_n(I_\asy,\sigma_\asy) := \int_{-\infty}^0  e^{nu}r_\text{asy}(u)du,\quad	\mathcal{J}_n(I_\asy,\sigma_\asy) := \int_{-\infty}^0  e^{nu}i_\text{asy}(u)du,
\end{equation*}
in terms of which
\begin{align*}
    & \int_0^{\infty} K(u)r_\asy(-u)du = \mathcal{Q}_0(I_\asy,\sigma_\asy)-\mathcal{Q}_1(I_\asy,\sigma_\asy), \\
    & \int_0^{\infty} K'(u)r_\asy(-u)du = \mathcal{Q}_1(I_\asy,\sigma_\asy).
\end{align*}
Observe that we can explicitly calculate 
\begin{equation*}
\mathcal{J}_n(I_\asy,\sigma_\asy)  = \frac{I_\asy}{2\sqrt{\pi}}U\left( \frac{n^2\sigma_\asy^2}{2} \right),
\end{equation*}
for all $n\geq 0$ where we recall $U(z):=\sqrt{\pi}e^{z}\erfc(\sqrt{z})$. Moreover, multiplying \eqref{eq:r_asy_eq_ode_0} by $ e^{nu}$ and integrating over $-\infty <u<0$ gives 
\begin{equation*}
    \mathcal{Q}_n = \frac{(1-m-w)\mathcal{J}_n + c_\asy r_\text{asy}(0)}{c_\asy n + m + \Ro I_\text{asy}} + \frac{\Ro I_\text{asy}}{c_\asy n + m + \Ro I_\text{asy}}\mathcal{Q}_{n+1},
\end{equation*}
where $r_\asy(0)$ is explicitly given in Remark \ref{rem:ode-for-rho}. Solving this recursion relation then gives
\begin{equation}\label{eq:Q-series}
    \mathcal{Q}_n(I_\asy,\sigma_\asy) = \sum_{l=n}^\infty  \frac{(1-m-w)\mathcal{J}_l + c_\asy r_\text{asy}(0)}{\prod_{k=n}^l (c_\asy k + m + \Ro I_\text{asy})} (\Ro I_\text{asy})^{l-n}.
\end{equation}
Note in particular that to evaluate $\mathcal{Q}_n$ only the values of $\mathcal{J}_l$ for $l\geq n$ are needed.  

In terms of $\mathcal{Q}_0$ and $\mathcal{Q}_1$ the nonlinear system \eqref{eq:nonlin_sys_0} becomes
\begin{subequations}\label{eq:nonlin_sys_2}
\begin{align}
    & \mathcal{Q}_0(I_\text{asy},\sigma_\asy) - \mathcal{Q}_1(I_\text{asy},\sigma_\asy) + \frac{m+ wI_\text{asy}}{m+\Ro I_\text{asy}} - \frac{1}{\Ro} = 0, \label{eq:nonlin_sys_2_1}\\
    & \sigma_\asy^2\Ro \mathcal{Q}_1(I_\text{asy},\sigma_\asy) - c_\asy = 0. \label{eq:nonlin_sys_2_2}
\end{align}
\end{subequations}
Since $0<I_\text{asy}<1-\Ro^{-1}$ we solve \eqref{eq:nonlin_sys_2_1} using a bisection method to get $I_\text{asy}$ in terms of $\sigma_\asy^2$ and then solve \eqref{eq:nonlin_sys_2_2} for $\sigma_\asy^2$ (specifically we used the \texttt{brentq} and \texttt{fsolve} routines from the \texttt{SciPy} library \cite{scipy}). Once $I_\asy$ and $\sigma_\asy$ are found we obtain an asymptotic approximation for $i_\asy(u)$ by \eqref{eq:i_e_gaussian} and for $r_\asy(u)$ by using \eqref{eq:r_asy_pos} for $u\geq 0$ and by solving \eqref{eq:r_asy_neg} for $u<0$.

We next collect explicit expressions for the functions $\mathcal{K}(\cdot)$ and $\mathcal{H}(\cdot)$ given by \eqref{eq:script-K-H-def} arising in the $\ep^2\ll m\ll \ep$ regime considered in  \S\ref{app:asymptotics-small-m}. Letting $u = -\log(z/\zeta)$ we obtain
\begin{subequations}
\begin{align}
    & \mathcal{K}(\zeta) = e^\zeta\int_0^\infty\exp\left(-\zeta(u+ e^{-u})\right)du = \frac{e^{\zeta}}{\zeta^{\zeta}}\gamma(\zeta,\zeta), \\ 
    & \mathcal{H}(\zeta) = e^\zeta\int_0^\infty\exp\left(-u-\zeta(u+ e^{-u})\right)du = \frac{e^\zeta}{\zeta^\zeta}\gamma(\zeta,\zeta)-\frac{1}{\zeta},
\end{align}
\end{subequations}
where $\gamma(s,x):=\int_0^x t^{s-1}e^{-t}dt$ is the lower incomplete gamma function \cite{NIST:DLMF}.

\section*{Acknowledgments}
We thank Yoichiro Mori for detailed interactions throughout the development of this work
We acknowledge support from the Simons Foundation Math+X grant to the University of
Pennsylvania.
\bibliographystyle{siamplain}
\bibliography{references}

\end{document}